\documentclass[prc,superscriptaddress,unsortedaddress,twocolumn,preprintnumbers,nofootinbib]{revtex4}

\usepackage{amsmath, amssymb}
\usepackage{url}
\usepackage{here}
\def\la{\mathrel{\mathpalette\fun <}}
\def\ga{\mathrel{\mathpalette\fun >}}
\def\fun#1#2{\lower3.6pt\vbox{\baselineskip0pt\lineskip.9pt
\def\dfrac#1#2{{\displaystyle\frac{#1}{#2}}}
\ialign{$\mathsurround=0pt#1\hfil##\hfil$\crcr#2\crcr\sim\crcr}}}
\newcommand{\beq}{\begin{equation}}
\newcommand{\eeq}{\end{equation}}
\newcommand{\bea}{\begin{eqnarray}}
\newcommand{\eea}{\end{eqnarray}}

\newcommand{\hyphen}{\mathchar`-}
\usepackage{graphicx}
\usepackage{times}
\usepackage{color}

\usepackage{ulem}
\usepackage{bm}

\renewcommand\sout{\bgroup\color{red} \ULdepth=-.5ex \ULset}

\newcommand{\comment}[1]{}

\begin{document}

\preprint{NITEP 91, YITP-21-15}

%---- title and author ----%
\title{Effect of deuteron breakup on the deuteron-$\Xi$ correlation function}

\author{Kazuyuki Ogata}
\email[]{kazuyuki@rcnp.osaka-u.ac.jp}
\affiliation{Research Center for Nuclear Physics, Osaka University, 
Ibaraki 567-0047, Japan}
\affiliation{Department, of Physics, Osaka City University, 
Osaka 558-8585, Japan}
\affiliation{Nambu Yoichiro Institute of Theoretical and Experimental Physics, 
Osaka City University, Osaka 558-8585, Japan}

\author{Tokuro Fukui}
\altaffiliation[Present address: ]{RIKEN Nishina Center, Wako 351-0198, Japan}
\affiliation{Yukawa Institute for Theoretical Physics, Kyoto University, Kyoto 606-8502, Japan}

\author{Yuki Kamiya}
\affiliation{CAS Key Laboratory of Theoretical Physics, Institute of Theoretical Physics, Chinese Academy of Sciences, Beijing 100190, China}

\author{Akira Ohnishi}
\affiliation{Yukawa Institute for Theoretical Physics, Kyoto University, Kyoto 606-8502, Japan}

\date{\today}

%---- abstract ----%
\begin{abstract}
\noindent
\textbf{Background:} The hadron-deuteron correlation function has attracted many interests as a potential method to access the three-hadron interactions. However, the weakly-bound nature of deuteron has not been considered in the preceding studies. \\
\textbf{Purpose:} The breakup effect of deuteron on the deuteron-$\Xi^-$ ($d$-$\Xi^-$) correlation function $C_{d\Xi^-}$ is investigated. \\
\textbf{Methods:} The $d$-$\Xi^-$ scattering is described by a nucleon-nucleon-$\Xi$ three-body reaction model. The continuum-discretized coupled-channels method, which is a fully quantum-mechanical and non-perturbative reaction model, is adopted. \\
\textbf{Results:} $C_{d\Xi^-}$ turns out to be sensitive to the strong interaction and enhanced by the deuteron breakup effect by 6--8~\% for the $d\hyphen\Xi^-$ relative momentum below about 70~MeV/$c$. Low-lying neutron-neutron continuum states are responsible for this enhancement.\\
\textbf{Conclusions:} Within the adopted model, the deuteron breakup effect on $C_{d\Xi^-}$ is found to be appreciable but not very significant. Except for the enhancement by several percent, studies on $C_{d\Xi^-}$ without the deuteron breakup effect can be justified.
\end{abstract}

\maketitle

\section{introduction}
\label{sec1}

Hadron-hadron ($hh$) interactions are the basic inputs
in describing hadronic many-body systems
such as nuclei, hadronic molecules, and nuclear matter.
The nucleon-nucleon ($NN$) interaction has been determined
by using $NN$ scattering data
and used in calculating various properties of nuclei and nuclear matter.
Other $hh$ pairs, by comparison,
scattering data are not enough to precisely determine
the interactions or not available,
so hypernuclear or mesic nuclear data have been invoked
to constrain $hh$ interactions such as the $\Xi$-hypernucleus~\cite{Xi-hypernucleus} for the $\Xi N$
interaction.

In these ten years, new techniques have been advanced
to elucidate the interactions of various $hh$ pairs.
Ab initio calculations of $hh$ interactions
based on lattice quantum chromodynamics (LQCD)~\cite{Luscher,HALQCD,HALQCD-pXi}
and the chiral effective field theory (chiral EFT)~\cite{chiEFT}
became available and some of the predictions have been examined to be reliable.
From an observational point of view,
femtoscopic studies of $hh$ interactions have been developed and advanced
recently~\cite{Lednicky,FemHHI,COSY-pL,STAR-pL,ALICE-pL-LL,STAR-LL,ALICE-pXi,ALICE-Nature,STAR-pW,ALICE-pK,ALICE-pSigma,ExHIC,Morita-LL,Ohnishi-LL-pK,Haidenbauer,Morita-pW,Hatsuda-pXi,Morita-pW-WW,Kamiya-pK,Mihaylov-CATS,Kamiya-pXi}.
The momentum correlation function of a particle pair
is defined as the two-particle production probability
normalized by the product of single-particle production probabilities
and is given by the convolution of the source function
and the squared relative wave function~\cite{KP}.
The correlation functions have been used to extract the source size
of stars and high-energy nuclear reactions
by assuming that the interaction between the particles are weak or well-known~\cite{HBT-GGLP}.
By comparison, when the source size is known,
one can use the correlation function to constrain the interaction between the particles~\cite{Lednicky,FemHHI}.
Actually, correlation functions have been measured recently
in high-energy nuclear collisions for various $hh$ pairs such as
$p\Lambda$~\cite{COSY-pL,STAR-pL,ALICE-pL-LL},
$\Lambda\Lambda$~\cite{STAR-LL,ALICE-pL-LL},
$p\Xi^-$~\cite{ALICE-pXi,ALICE-Nature},
$p\Omega$~\cite{STAR-pW,ALICE-Nature},
$pK^-$~\cite{ALICE-pK},
and $p\Sigma^0$~\cite{ALICE-pSigma},
and these data have been used to constrain the $hh$ interactions~\cite{Lednicky,FemHHI,ExHIC,Morita-LL,Ohnishi-LL-pK,Haidenbauer,Morita-pW,Hatsuda-pXi,Morita-pW-WW,Kamiya-pK,Mihaylov-CATS}.

As the next step in the femtoscopic studies of $hh$ interactions,
the hadron-deuteron ($hd$) correlation functions would be
promising as discussed in Refs.~\cite{MS20,Haidenbauer-dL,Etminan-dW}.
The $hd$ correlation function has several merits to study.
First, it is sensitive to the $hd$ scattering length,
which can be compared with the precise few-body calculation results.
Second, there is a possibility that one can access the hadron-nucleon-nucleon
three-body interaction, which would be important to evaluate
the dense matter equation of state~\cite{3BF-EOS}
and the three-body bound state if exists.
Third, by using the hadron-nucleus correlation function,
different spin-isospin components in the hadron-nucleon interaction may be resolved.
In the $\Lambda{d}$ correlation function, for example, 
the $s$-wave function contains the doublet ($^2\mathrm{S}_{1/2}$)
and quartet ($^4\mathrm{S}_{3/2}$) components.
Since the scattering length of the former (doublet channel) is strongly constrained
by the binding energy of the hypertriton ($^3_\Lambda\mathrm{H}$),
the $\Lambda{d}$ correlation function data will tell us the quartet channel
scattering length~\cite{Haidenbauer-dL}.
The scattering length in both channels are helpful
to resolve the $\Lambda{N}$ interactions in the spin-singlet and triplet channels
and to deduce the strength of the $\Lambda{NN}$ three-body force.
In order to extract these interesting ingredients from the $hd$ correlation functions,
precise theoretical estimates are necessary.
One of the important issues in the $hd$ correlation function
is the large size and the small binding energy of the deuteron.
In the previous exploratory theoretical studies of the $pd$~\cite{MS20} and
$hd$ correlation functions,
$K^{-}d$~\cite{MS20}, $\Lambda{d}$~\cite{Haidenbauer-dL}, and $\Omega^-{d}$~\cite{Etminan-dW},
the $hd$ interaction is evaluated using the intrinsic deuteron wave function,
but the deuteron breakup effects are ignored; in Ref.~\cite{MS20}, it has been conjectured that the deuteron breakup effect can be effectively taken into account by increasing the source size of the deuteron source function.
Besides, the asymptotic wave function is assumed even in the interaction range
by using the analytical Lednicky-Lyuboshitz formula~\cite{Lednicky}
in Refs.~\cite{MS20,Haidenbauer-dL}.
In order to take account of the deuteron compositeness
and the wave function inside the interaction range,
it is necessary to obtain the three-body wave function
with the deuteron breakup effects
under the boundary condition of $h$ and $d$ in the asymptotic region.

The purpose of this study is to predict the $d$-$\Xi^-$ correlation function with a $N+N+\Xi$ three-body model including the effects of the breakup states of deuteron. To achieve this, we adopt the continuum-discretized coupled-channels method (CDCC)~\cite{Kam86,Aus87,Yah12}. CDCC is one of the most accurate and flexible reaction models for describing processes in which a weakly-bound particle is involved. The theoretical foundation of CDCC is given in Refs.~\cite{Aus89,Aus96} in connection with the distorted-wave Faddeev theory~\cite{BR82}. This has been confirmed also numerically in Refs.~\cite{Del07,Upa12,OY16} on $d$-nucleus reactions. The validation of CDCC in a similar manner for $d$-$N$ scatterings has not been done mainly because of the difficulty in treating the antisymmetrization between each nucleon inside $d$ and the other nucleon outside $d$. Fortunately, however, such a complicated antisymmetrization of two nucleons is not needed in the $d$-$\Xi^-$ scattering. One can, therefore, expect the validation of CDCC confirmed so far also for the $d$-$\Xi^-$ scattering. Note that the antisymmetrization between two nucleons inside $d$ is included as shown in Sec.~\ref{sec2}. In CDCC, the wave function of the reaction system is described in terms of a finite number of channels. The Argonne V4' (AV4') nucleon-nucleon ($NN$) interaction~\cite{WP02} and the $N$-$\Xi$ interaction obtained by LQCD~\cite{HALQCD-pXi} are employed. Through the spin- and isospin-dependence of the $N$-$\Xi$ interaction, the total isospin ($T$) and spin ($S$) of the $NN$ system are not conserved. We include both of the $s$-wave channels, $(T,S)=(0,1)$ and $(T,S)=(1,0)$ states, in the present CDCC calculation.

As a first step of the three-body study on the $d$-$\Xi^-$ correlation function with CDCC, we make the following approximations. First, the Coulomb interaction between charges $+e$ and $-e$ is assumed to be present in all channels. Second, the orbital angular momentum between the two nucleons and that between $\Xi$ and the center-of-mass (c.m.) of the $NN$ system are both limited to zero. Third, a source function of $d$-$\Xi^-$ is considered rather than that of the $NN\Xi$.
Fourth, we ignore the isospin dependence of masses of $N$ and $\Xi$ baryons.
We discuss the properties of the $N+N+\Xi$ three-body system relevant to the $d$-$\Xi^-$ scattering under these conditions and clarify the $NN$ breakup effect on the $d$-$\Xi^-$ correlation function.

The construction of this paper is as follows. In Sec.~\ref{sec2}, we describe the formulation of the $d$-$\Xi^-$ correlation function based on CDCC. The numerical inputs are given in Sec.~\ref{sec31}. The calculated $d$-$\Xi^-$ correlation function and its convergence feature regarding the model space of CDCC are shown in Sec.~\ref{sec32}. The dependence of the correlation function on the source size of the source function is also discussed. In Sec.~\ref{sec33}, properties of the $NN$ breakup states included in the CDCC calculation are shown, and those of the coupling potentials of the $NN$-$\Xi$ system are investigated in Sec.~\ref{sec34}. The resulting $NN$-$\Xi$ scattering wave functions are discussed in Sec.~\ref{sec35}. Finally, a summary is given in Sec.~\ref{sec4}.

\section{formalism}
\label{sec2}

The discretized continuum states of the $NN$ system in CDCC are given by
\begin{equation}
\varphi_{iTS}\left(  r\right)  =\frac{1}{\sqrt{\Delta_{iTS}}}\int_{k_{iTS}}%
^{k_{iTS}+\Delta_{iTS}}\varphi_{TS}\left(  k,r\right)  dk,
\label{av}%
\end{equation}%
where $i$, $T$, and $S$ are the energy index, the total isospin, and the total spin of the $NN$ system, respectively. $r$ is the distance between the two nucleons and $k$ is their relative wave number. The $NN$ orbital angular momentum is restricted to 0 in this study; because of the antisymmetrization condition of the $NN$ system, we only include the states with $S+T=1$. $\varphi_{TS}$ is the $NN$ scattering wave function satisfying
\begin{equation}
\left[  -\frac{\hbar^{2}}{2\mu_{r}}\frac{d^{2}}{dr^{2}}+V^{(NN)}_{TS}\left(
r\right)  \right]  \varphi_{TS}\left( k,r\right)  =\varepsilon%
\varphi_{TS}\left( k,r\right),
\label{seqnn}
\end{equation}
where $\mu_{r}$ is the $NN$ reduced mass, $V_{TS}^{(NN)}$ is the $NN$ interaction of the central type, and $\varepsilon=\hbar^2 k^2/(2\mu_r)$. $\varphi_{TS}$ is solved under the following boundary condition
\begin{equation}
\varphi_{TS}\left(  k,r\right) \rightarrow
\sqrt{\frac{2}{\pi}} \sin\left[ kr + \delta^{(NN)}_{TS}(k) \right], \quad (r \to \infty)
\end{equation}
with $\delta^{(NN)}_{TS}$ being the $NN$ scattering phase shift in the $s$-wave. As shown in Eq.~(\ref{av}), $\varphi_{TS}$ is averaged over $k$ within the bin of $k$ characterized by the lower limit $k_{iTS}$ and the width $\Delta_{iTS}$, which is called a {\lq\lq}momentum bin'' or {\lq\lq}bin state'' in convention. The eigenenergy $\varepsilon_{iTS}$ of $\varphi_{iTS}$ is defined by
\begin{eqnarray}
\varepsilon_{iTS}
&=&
\frac{1}{\Delta_{iTS}}\int_{k_{iTS}}^{k_{iTS}+\Delta_{iTS}%
}\frac{\hbar^{2}k^2}{2\mu_{r}}dk
\nonumber \\
&=&
\frac{\hbar^2}{2\mu_r}
\left(k_{iTS}^2+k_{iTS} \Delta_{iTS}+\frac{\Delta_{iTS}^2}{3}
\right).
\label{epsi}
\end{eqnarray}
In what follows, for the simple notation, we use the channel index $c$ that represents $(i,T,S)$ altogether; $c=0$ corresponds to the deuteron ground state. The discretized continuum states $\varphi_c$ are orthonormal
\begin{equation}
\int
\varphi^*_{c'}\left(  r\right) \varphi_{c}\left(  r\right) dr
=\delta_{c'c}
\end{equation}
and satisfy
\begin{equation}
\left[  -\frac{\hbar^{2}}{2\mu_{r}}\frac{d^{2}}{dr^{2}}+V^{(NN)}_{TS}\left(
r\right)  \right]  \varphi_{c}\left(r\right)  =\varepsilon_c
\varphi_{c}\left(r\right).
\label{seqnn2}
\end{equation}

The $s$-wave component of the total ($NN\Xi$) wave function that satisfies the outgoing boundary condition is given by
\begin{eqnarray}
\Psi_{M_0 \mu_0}^{(+)}(r,R)
&=&
\sqrt{4\pi}\sum_{\sigma m_{\sigma}}
\left(1 M_0 \frac{1}{2} \mu_0 \Big|\sigma m_{\sigma}\right)
e^{i\sigma_{0}} \nonumber\\
& & \times\sum_{c'}
\frac{\varphi_{c'}\left(r\right)}{r}
\frac{\chi_{c'}^{(\sigma)}\left( K_{c'},R\right)}{K_0R}
\frac{1}{4\pi}  \nonumber\\
& & \times \Upsilon_{S'}^{\left(\sigma m_\sigma\right)}
\Theta_{T'}^{\left(\frac{1}{2},-\frac{1}{2} \right)},
\label{totwf}%
\end{eqnarray}
where $R$ is the distance between $\Xi$ and the c.m. of the $NN$ system, $(abcd|ef)$ is the Clebsh-Gordan coefficient, and $\sigma_0$ is the $s$-wave Coulomb phase shift. $M_0$ and $\mu_0$ represent the third component of the spin of $d$ and that of $\Xi^-$, respectively, in the incident channel. $\sigma$ is the channel spin and $m_\sigma$ is its third component. Note that the channel isospin and its third component are fixed at $1/2$ and $-1/2$, respectively, in the $d$-$\Xi^-$ scattering; isospin $3/2$ channels do not couple with the $d\hyphen\Xi^-$ channel.
The channel-spin wave function is defined by
\begin{equation}
\Upsilon_S^{\left(\sigma m_\sigma\right)}
=
\left[
\eta_{S}^{\left(NN\right)}
\otimes
\eta_{\frac{1}{2}}^{\left(\Xi\right)}
\right] _{\sigma m_{\sigma}}
\end{equation}
with $\eta_S^{(NN)}$ ($\eta_{1/2}^{(\Xi)}$) being the spin wave function of the $NN$ system ($\Xi$). Similarly, the channel-isospin wave function is given by
\begin{equation}
\Theta_{T}^{\left(\frac{1}{2},-\frac{1}{2} \right)}=
\left[  \zeta_{T}^{\left(  NN\right)  }
\otimes
\zeta_{1/2}^{\left(  \Xi\right)  }
\right] _{\frac{1}{2},-\frac{1}{2}},
\end{equation}
where $\zeta_T^{(NN)}$ and $\zeta_{1/2}^{(\Xi)}$ are the isospin wave functions of the $NN$ system and $\Xi$, respectively.

The wave number of $\Xi$ relative to the c.m. of the $NN$ system in channel $c$, denoted by $K_c$, is determined by the conservation of the total energy $E_{\rm tot}$:
\begin{equation}
\frac{\hbar^2 K_c^2}{2\mu_R}+\varepsilon_c=E_{\rm tot},
\end{equation}
where $\mu_R$ is the reduced mass between $\Xi$ and the $NN$ system.
$\chi_{c}^{(\sigma)}$ is the radial part of the $NN$-$\Xi$ scattering wave function in channel $c$ multiplied by $K_0 R$. Its boundary condition outside the strong interaction range is given by
\begin{equation}
\chi_{c}^{(\sigma)}( K_{c},R)
\rightarrow
\frac{i}{2}\left[  {\cal U}_{0,\eta_c}^{(-)}(K_{c}R)  \delta_{c0}
-\sqrt{\frac{K_{0}}{K_{c}}}
S_{c}^{(\sigma)}{\cal U}_{0,\eta_c}^{(+)}( K_{c}R)  \right]
\label{bco}
\end{equation}
for $K_c^2>0$ (open channels) and by
\begin{equation}
\chi_{c}^{(\sigma)}(K_{c},R)
\rightarrow
-\frac{i}{2}S_{c}^{\left(
\sigma \right)  }W_{-\eta_{c},1/2}(  -2iK_{c}R)
\label{bcc}
\end{equation}
for $K_c^2<0$ (closed channels). Here, $S_{c}^{\left(\sigma \right)}$ is the scattering matrix ($S$ matrix) and ${\cal U}_{0,\eta_c}^{(+)}$ (${\cal U}_{0,\eta_c}^{(-)}$) is the $s$-wave outgoing (incoming) Coulomb wave function and $W_{-\eta_{c},1/2}$ is the $s$-wave Whittaker function with $\eta_{c}$ being the Sommerfeld parameter
\begin{equation}
\eta_{c}=-\frac{\mu_R e^2}{\hbar^2 K_c}.
\end{equation}
The $S$ matrix has the following unitarity condition:
\begin{equation}
\sum_{c \in {\rm open}\mbox{ }{\rm channels}}\left| S_{c}^{\left(\sigma \right)} \right|^2 = 1.
\end{equation}

$\Psi_{M_0\mu_0}$ satisfies the Sch\"odinger equation
\begin{equation}
\left[H-E_{\rm tot}\right]\Psi_{M_0\mu_0}^{(+)}(r,R)=0
\label{scheq}
\end{equation}
with
\begin{equation}
H\equiv
-\frac{\hbar^2}{2\mu_R}{\bm \nabla}^2_R
+\sum_{i=1,2} V^{(N\Xi)}(R_i)
+V^{\rm C}(R)+h_{NN},
\label{htot}
\end{equation}
where $R_1=|{\bm R}-{\bm r}/2|$ and $R_2=|{\bm R}+{\bm r}/2|$
are the distances between $\Xi$ and one of the nucleons, $V^{\rm C}$ is the Coulomb interaction between the charges $+e$ and $-e$ at a distance of $R$, and $h_{NN}$ is the $NN$ internal Hamiltonian defined by
\begin{equation}
h_{NN}=-\frac{\hbar^2}{2\mu_r}{\bm \nabla}^2_r + \sum_{TS}V^{(NN)}_{TS}(r) {\cal P}^{(NN)}_{TS}.
\end{equation}
Here and in what follows, ${\cal P}^{NX}_{\alpha\beta}$ represents the projection operator onto the isospin $\alpha$ and spin $\beta$ state of the $NX$ system. The $N\Xi$ interaction is given by
\begin{equation}
V^{(N\Xi)}(R_i)=\sum_{ts} V_{ts}^{(N\Xi)}(R_i) {\cal P}^{(N\Xi)}_{ts}.
\end{equation}

One obtains the following coupled-channel (CC) equations by
inserting Eqs.~(\ref{totwf}) and (\ref{htot}) into Eq.~(\ref{scheq}), multiplying the equation by
\begin{equation}
\frac{\varphi^*_{c}\left(r\right)}{r}
\frac{1}{4\pi} \Upsilon_{S}^{\left(\sigma m_\sigma\right)*}
\Theta_{T}^{\left(\frac{1}{2},-\frac{1}{2} \right)*}
\end{equation}
from the left, and making integration over coordinates other than $R$; $NN$ relative coordinate ${\bm r}$, the internal coordinates associated with the spin and isospin, and the solid angle $\Omega_R$ of ${\bm R}$,
\begin{align}
\left[
-\frac{\hbar^{2}}{2\mu_{R}}\frac{d^{2}}{dR^{2}}+V^{\rm C}\left(R\right)-E_{c}
\right] &
\chi_{c}^{\left(\sigma\right)}( K_{c},R)
\nonumber \\
=-\sum_{c'} & U_{cc'}^{\left(\sigma\right)}\left(R\right)
 \chi_{c'}^{\left(\sigma\right)}\left(K_{c'},R\right)
\label{cceq}
\end{align}
with
\begin{equation}
E_c = E_{\rm tot} - \varepsilon_c.
\end{equation}
The coupling potentials $U_{cc'}^{\left(\sigma\right)}$ are given by
\begin{equation}
U_{cc'}^{\left(\sigma\right)}\left(R\right)
=2\sum_{ts}
w_{tTT'}^{(1/2)}
w_{sSS'}^{(\sigma)}
f^{(ts)}_{cc'}(R),
\label{ccpot}
\end{equation}
where
\begin{equation}
f^{(ts)}_{cc'}(R)\equiv
\int\varphi_{c}^*(r)
V_{ts;0}^{(N\Xi)}(R,r)
\varphi_{c'}(r)
dr,
\end{equation}
\begin{eqnarray}
w_{aBC}^{(\alpha )}
&\equiv&
(2a+1)\sqrt{2B+1}\sqrt{2C+1}
\nonumber \\
& & \times
W(1/2,1/2,1/2,\alpha;aB)
W(1/2,1/2,1/2,\alpha;aC)
\nonumber \\
\end{eqnarray}
with $W(j_1 j_2 j_3 j_4;j_5j_6)$ being the Racah coefficient,
and the monopole component of the $N\Xi$ potential is given by
\begin{equation}
V_{ts;0}^{(N\Xi)}(R,r)
=
\frac{1}{2}\int_{-1}^{1}
V_{ts}^{(N\Xi)}\big(
\sqrt{R^2+r^2/4-Rrx}\,
\big)
dx.
\end{equation}
By putting explicit values of the Racah coefficient, one finds
\begin{eqnarray}
U_{i01,i'01}^{\left( 1/2 \right)}\left(R\right)
&=&
\frac{1}{8}\left[
 3f^{(00)}_{i01,i'01}(R)
+9f^{(10)}_{i01,i'01}(R)\right. \nonumber \\
&&
\left.
+ f^{(01)}_{i01,i'01}(R)
+3f^{(11)}_{i01,i'01}(R)
\right],
\label{ucctete}
\end{eqnarray}
\begin{eqnarray}
U_{i01,i'10}^{\left( 1/2 \right)}\left(R\right)
&=&
\frac{1}{8}\left[
  f^{(00)}_{i01,i'10}(R)
-3f^{(10)}_{i01,i'10}(R)\right. \nonumber \\
&&
\left.
-3f^{(01)}_{i01,i'10}(R)
+3f^{(11)}_{i01,i'10}(R)
\right] \nonumber \\
&=&
U_{i10,i'01}^{\left( 1/2 \right)}\left(R\right),
\label{ucctese}
\end{eqnarray}
\begin{eqnarray}
U_{i10,i'10}^{\left( 1/2 \right)}\left(R\right)
&=&
\frac{1}{8}\left[
 3f^{(00)}_{i10,i'10}(R)
+ f^{(10)}_{i10,i'10}(R)\right. \nonumber \\
&&
\left.
+9f^{(01)}_{i10,i'10}(R)
+3f^{(11)}_{i10,i'10}(R)
\right],
\label{uccsese}
\end{eqnarray}
\begin{equation}
U_{i01,i'01}^{\left( 3/2 \right)}\left(R\right)
=
\frac{1}{2}\left[
  f^{(01)}_{i01,i'01}(R)
+3f^{(11)}_{i01,i'01}(R)
\right],
\label{ucc3h}
\end{equation}
\begin{equation}
U_{i01,i'10}^{\left( 3/2 \right)}\left(R\right)
=U_{i10,i'01}^{\left( 3/2 \right)}\left(R\right)
=U_{i11,i'11}^{\left( 3/2 \right)}\left(R\right)
=0.
\end{equation}

Under the assumption that only the $s$-wave component is affected by the strong interaction, the total wave function of the reaction system having the incoming boundary condition is expressed by
\begin{equation}
\Psi_{M_0\mu_0}^{(-){\rm tot}}(r,{\bm R})
=
\Psi_{M_0\mu_0}^{(-)}(r,R)
+
\psi_{M_0\mu_0}^{{\rm C}(-)}(r,{\bm R}).
\label{psimin}
\end{equation}
Here, $\Psi_{M_0\mu_0}^{(-)}$ is the time-reversal of $\Psi_{M_0\mu_0}^{(+)}$, the explicit form of which is given in Appendix~\ref{a1}, and
\begin{eqnarray}
\psi_{M_0\mu_0}^{{\rm C}(-)}(r,{\bm R})
&\equiv&
\frac{\varphi_0 (r)}{r}\frac{1}{\sqrt{4\pi}}
\left[\phi_{\bm K}^{{\rm C}(-)}({\bm R})-\frac{e^{-i\sigma_0}F_0(K_0R)}{K_0R}\right]
\nonumber \\
& &
\times \eta_{1M_0}^{\left(NN\right)}
\eta_{\frac{1}{2} \mu_0}^{\left(\Xi\right)}
\zeta_{00}^{\left(NN\right)}
\zeta_{\frac{1}{2},-\frac{1}{2}}^{\left(\Xi\right)}
\label{psicmin}
\end{eqnarray}
with $\phi^{{\rm C}(-)}$ being the Coulomb scattering wave function with the incoming boundary condition and $F_0$ the $s$-wave Coulomb wave function that is regular at the origin.

We follow Ref.~\cite{KP} for the calculation of the $d$-$\Xi^-$ correlation function $C_{d\Xi^-}$. To implement the three-body scattering wave function $\Psi_{M_0\mu_0}^{(-){\rm tot}}$ of Eq.~(\ref{psimin}) into $C_{d\Xi^-}$, we first take its overlap with
\begin{equation}
\Phi_{cM\mu\nu_{T}\nu}(r)
=
\frac{\varphi_{c}(r)}{r}\frac{1}{\sqrt{4\pi}}%
\eta_{SM}^{\left(NN\right)}
\eta_{\frac{1}{2} \mu}^{\left(\Xi\right)}
\zeta_{T \nu_T}^{\left(NN\right)}
\zeta_{\frac{1}{2}\nu}^{\left(\Xi\right)}.
\end{equation}
We then take a summation over $c$ and obtain
\begin{align}
C_{d\Xi^-}(K_0)
=&
4\pi\int R^2 dR\, {\cal S}(R) 
\sum_{L=1}
(2L+1) \left[ \frac{F_L(K_0R)}{K_0R}\right]^2
\nonumber \\
+&\frac{2\pi}{3}\int R^2 dR\, {\cal S}(R) \sum_{c\sigma} (2\sigma+1)
\left|
\frac{\chi_{c}^{(\sigma)}(K_{c},R)}{K_0R}
\right|^2, \nonumber \\
\label{corr2}
\end{align}
where ${\cal S}$ is the source function of the $d\hyphen\Xi^-$ pair. We have assumed that ${\cal S}$ does not depend on ${\bm r}$; the channel dependence of ${\cal S}$ is also disregarded for simplicity. $F_L$ is the same as $F_0$ in Eq.~(\ref{psicmin}) but for an orbital angular momentum $L$.

It should be noted that, because we deal with the three-body wave function having the incoming boundary condition, $c\neq 0$ channels correspond to the processes in which {\it initially} three-particles ($N+N+\Xi$) exist and through the propagation the transition to the $c=0$ channel occurs. Then, eventually, the $d$-$\Xi^-$ two-particle state with the relative momentum $\hbar c K_0$ is observed.

While we consider the $d\hyphen\Xi^-$ source function,
it is, in principle, possible to start from the $NN\Xi$ source function 
and to evaluate the deuteron formation dynamically
by using the $NN$ relative wave function $\varphi_c(r)$.
This process was discussed in detail in Ref. [29]. When the three-body source function for the $NN\Xi \to d\Xi^-$ process is considered and the center-of-mass and deuteron intrinsic coordinates are integrated out, the source function in the relative coordinate of $d$-$\Xi^-$ was found to be $D_{3r}(\bm{R}) \propto \exp[-R^2/(3R_s^2)]$ with $R_s$ being the single hadron source size [29]. By comparing it with the $d$-$\Xi^-$ source function adopted in the present work, $\mathcal{S}(R)\propto \exp[-R^2/(4b^2)]$, it it found that the size parameter needs to be taken as $b \simeq \sqrt{3/4}R_s$. Thus, we need to take care of the difference of $b$ and the single hadron source size $R_s$. The combined treatment of the pre-formed deuteron source function and the three-body source function is a theoretical challenge, beyond the scope of this paper, and left as a future work.
Results with this extension will be reported elsewhere.

\section{results and discussion}
\label{sec3}

\subsection{Numerical inputs}
\label{sec31}

We adopt the Argonne V4' parameter~\cite{WP02} for the $NN$ interaction. The triplet-even $^{13}{\rm S}_{1}$ and the singlet-even $^{31}{\rm S}_{0}$ states are taken into account. The continua of these states are truncated at $k_{\rm max}=2.0$~fm$^{-1}$ ($\sim 400$~MeV/$c$); the size $\Delta_c$ of the bin state is set to 0.2~fm$^{-1}$ ($\sim 40$~MeV/$c$) and 0.005~fm$^{-1}$ ($\sim 1$~MeV/$c$) for the $^{13}{\rm S}_{1}$ and $^{31}{\rm S}_{0}$ states, respectively\footnote{Because the breakup states are characterized by wave numbers, not momenta, in the CDCC code employed, we represent $k_{\rm max}$ and $\Delta_c$ in the unit of fm$^{-1}$.}. $r_{\rm max}=20$~fm is taken for evaluating the folded potentials.

As for the $N$-$\Xi$ strong interaction, we employ the parameterization by the LQCD work at $a/t=11$~\cite{HALQCD-pXi}. In the original parameterization, the $N$-$\Xi$ interaction $V_{ts}^{(N\Xi)}$ for each spin ($s$) and isospin ($t$) channel was expressed by the sum of one Yukawa function (with a form factor), one squared Yukawa, and three Gaussians. In this study, we expand each of the former two by 30 Gaussians; the range parameters are chosen in a geometric progression and the minimum and maximum ranges are optimized for each $st$ channel. It is found that $V_{ts}^{(N\Xi)}$ thus obtained gives the $N$-$\Xi$ phase shift that agrees with the result with the original $V_{ts}^{(N\Xi)}$ for six digits. By expressing all the terms of $V_{ts}^{(N\Xi)}$ by Gaussians, one can use the simple analytic form of Eq.~(\ref{monog}) for the monopole component of the $N$-$\Xi$ interaction.

The CC equations (\ref{cceq}) are integrated up to $R=10$~fm. The Coulomb interaction $V^{\rm C}$ is taken to be
\begin{equation}
V^{\rm C}(R)=
\left\{
\begin{array}
[c]{lc}%
\displaystyle\frac{-e^2}{2R_0}
\left(3-\displaystyle\frac{R^2}{R_0^2}\right) & \quad (R \le R_0) \\
\displaystyle\frac{-e^2}{R} & \quad  (R > R_0)
\end{array}
\right.,
\end{equation}
with $R_0=1.5$~fm. The dependence of the numerical results shown below on $R_0$ is found to be negligibly small (less than 1\%).

The source function ${\cal S}$ is assumed to have a Gaussian form
\begin{equation}
{\cal S}(R)=\frac{1}{(4\pi b^2)^{3/2}}e^{-R^2/(4b^2)}.
\end{equation}
The source size $b$ of the source function is taken to be 1.2~fm; in Fig.~\ref{fig4}, results with $b=1.6$ and 3.0~fm are shown for comparison.
In the evaluation of the correlation function, the integration over $R$ is carried out up to $R_{\rm max}=10$~fm (15~fm) when $b=1.2$~fm and 1.6~fm (3.0~fm), and the maximum $L$ is taken to be a larger of $K_0 R_{\rm max}$ and 5.

\subsection{Correlation function}
\label{sec32}

%---- figure 1 ----%
\begin{figure}[H]
\begin{center}
\includegraphics[width=0.48\textwidth]{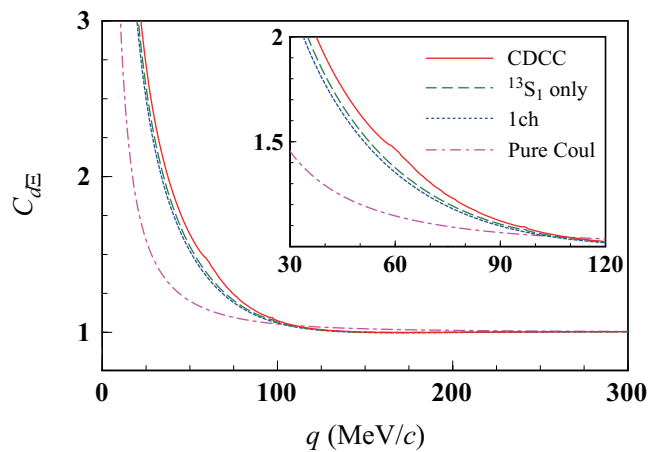}
\caption{$d$-$\Xi^{-}$ correlation function as a function of the relative momentum $q$. The solid, dashed, dotted, and dash-dotted lines represent the result of CDCC, that with the $^{13}{\rm S}_1$ breakup states only, the result of the single-channel calculation (without breakup states), and the result with switching the strong interactions off, respectively. The inset is an enlarged result for $30~{\rm MeV}/c \le q \le 120~{\rm MeV}/c$.}
\label{fig1}
\end{center}
\end{figure}
We show in Fig.~\ref{fig1} $C_{d\Xi^-}$ as a function of $q\equiv\hbar c K_0$. The inset is an enlarged figure in the region of $30~{\rm MeV}/c \le q \le 120~{\rm MeV}/c$. The solid (red) line represents the result calculated with the present framework of CDCC. The dotted (blue) line is the result of the single-channel calculation, that is, only the ground state of deuteron is considered. If we take only the $^{13}{\rm S}_1$ channels in $NN$ into account, the dashed (green) line is obtained. The dash-dotted (purple) line is the result obtained with all the strong interactions turned off. For a simple notation, below we designate the $^{13}{\rm S}_1$ ($^{31}{\rm S}_0$) channel as the $pn$ ($nn$) channel.

The solid line shows a clear enhancement relative to the dash-dotted line for $q \le 100~{\rm MeV}/c$, which indicates that the correlation due to the strong interaction can be deduced from $C_{d\Xi^-}$.
The difference of the solid line from the dotted line represents an increase in $C_{d\Xi^-}$ by the deuteron breakup effect, which is about 6--8~\% for $30~{\rm MeV}/c \le q \le 70~{\rm MeV}/c$. At larger $q$, the enhancement due to deuteron breakup decreases monotonically and becomes less than 1\% for $q>100~{\rm MeV}/c$. We discuss the deuteron breakup effect in more detail in Sec.~\ref{sec35}. The small difference between the dashed and dotted line indicates that the $nn$ breakup states are more significant than the $pn$ breakup states. This can be understood by the behavior of the CC potentials as discussed in Sec.~\ref{sec34}. With a closer look, a shoulder structure is found in the solid line at around 60~MeV$/c$. This corresponds to the strong coupling to low-lying $nn$ breakup states located just below the scattering threshold; the channel energy $E_c$ is negative and close to 0. We will return to this point soon below and in Sec.~\ref{sec35}. Compared with the net effect of the strong interaction (difference between the solid and dash-dotted lines), the deuteron breakup effect is found to be not very significant. In other words, including only the deuteron ground state in the calculation of $C_{d\Xi^-}$ will be useful except that it will miss a further increase in the correlation function by several percent below about 70~MeV$/c$.

%---- figure 2 ----%
\begin{figure}[H]
\begin{center}
\includegraphics[width=0.48\textwidth]{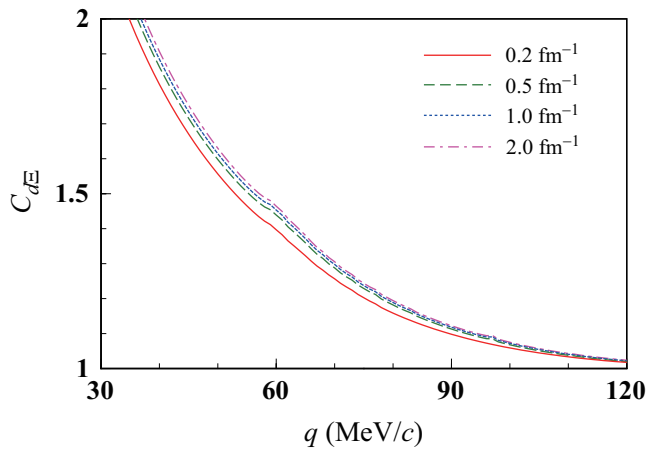}
\caption{Convergence of the $d$-$\Xi$ correlation function regarding $k_{\rm max}$. The horizontal axis is the $d\hyphen\Xi^-$ relative momentum. The solid, dashed, dotted, and dash-dotted lines correspond to $k_{\rm max}=0.2$, 0.5, 1.0, and 2.0~fm$^{-1}$, respectively.}
\label{fig2}
\end{center}
\end{figure}
Figure~\ref{fig2} displays the convergence of $C_{d\Xi^-}$ regarding $k_{\rm max}$. In all the calculations, we take the size $\Delta_c$ of the bin state to be 0.2 fm$^{-1}$ (0.005 fm$^{-1}$) for the $pn$ ($nn$) continuum. The solid (red), dashed (green), dotted (blue), and dash-dotted (purple) lines correspond to $k_{\rm max}=0.2$, 0.5, 1.0, and 2.0~fm$^{-1}$, respectively. The dash-dotted line is the same as the solid line in Fig.~\ref{fig1}. The result with $k_{\rm max}=2.5$~fm$^{-1}$ is found to agree with the dash-dotted line within the width of the line (not shown). It should be noted that almost all of the $NN$ states included in the converged CDCC calculation serve as a closed channel. For instance, at $q=100$~MeV$/c$ ($E_0\sim 4.22$~MeV), $NN$ states having $k\ga 0.32$~fm$^{-1}$ are all closed, whereas we need the $NN$ states up to 2.0~fm$^{-1}$ ($\varepsilon\sim 166$~MeV) to achieve a convergence of $C_{d\Xi^-}$.

%---- figure 3 ----%
\begin{figure}[H]
\begin{center}
\includegraphics[width=0.48\textwidth]{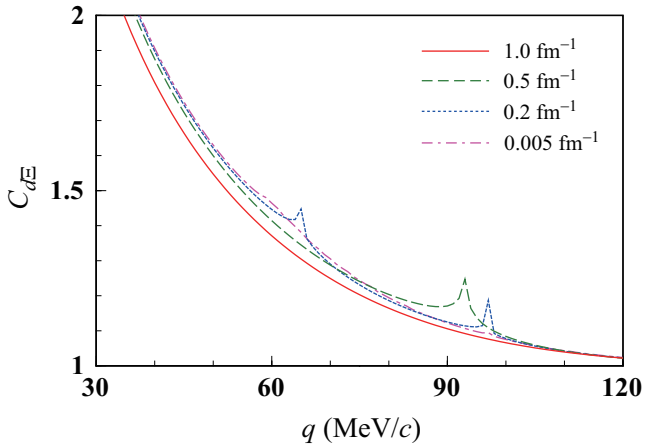}
\caption{Convergence of the $d$-$\Xi$ correlation function regarding $\Delta_c$ for the $nn$ continuum; $\Delta_c$ for the $pn$ continuum is taken to be 0.2~fm$^{-1}$. The horizontal axis is the $d\hyphen\Xi^-$ relative momentum. The solid, dashed, dotted, and dash-dotted lines correspond to $\Delta_c=1.0$, 0.5, 0.2, and 0.005~fm$^{-1}$, respectively.}
\label{fig3}
\end{center}
\end{figure}
%
%---- figure 4 ----%
\begin{figure}[t]
\begin{center}
\includegraphics[width=0.48\textwidth]{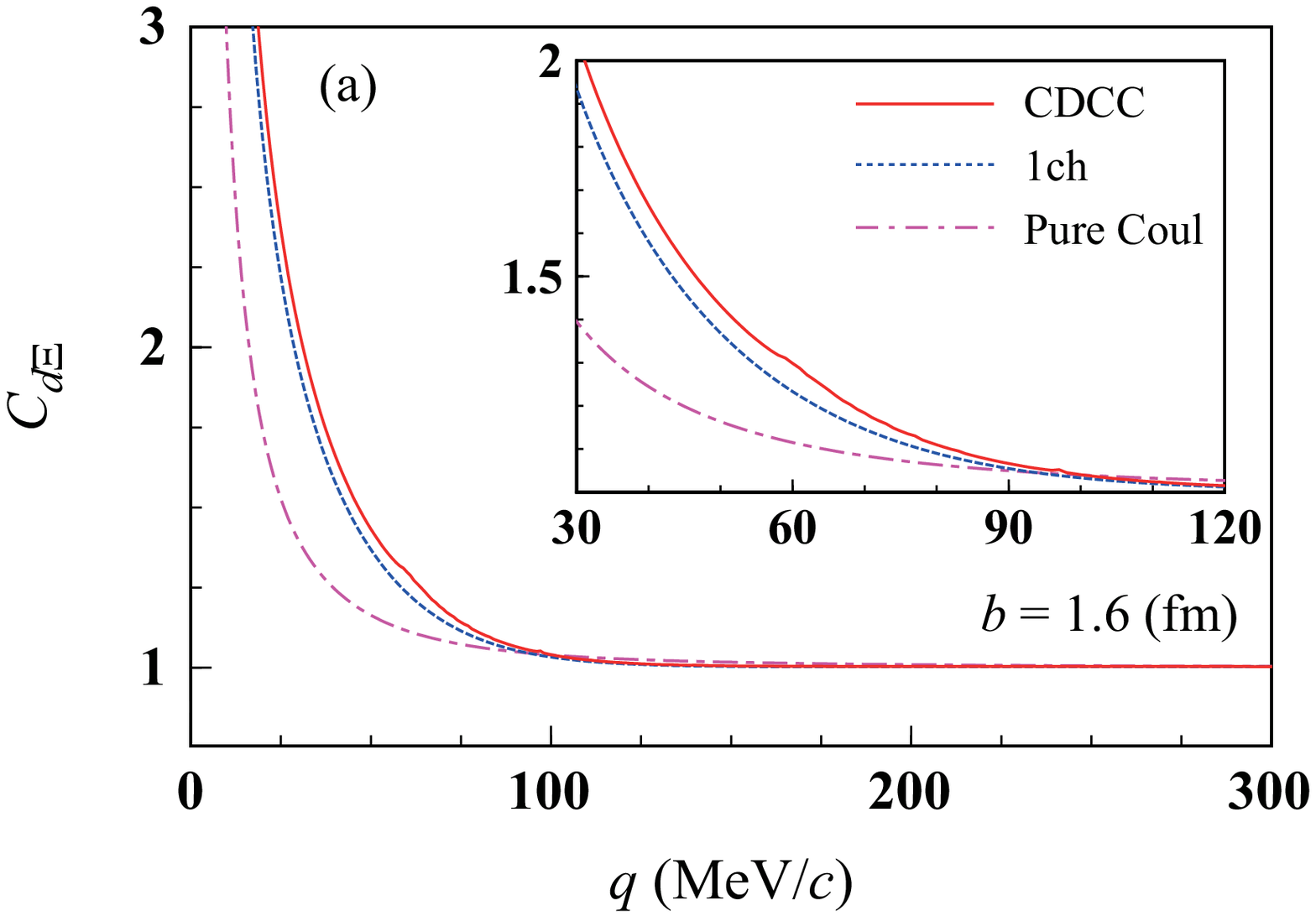}
\includegraphics[width=0.48\textwidth]{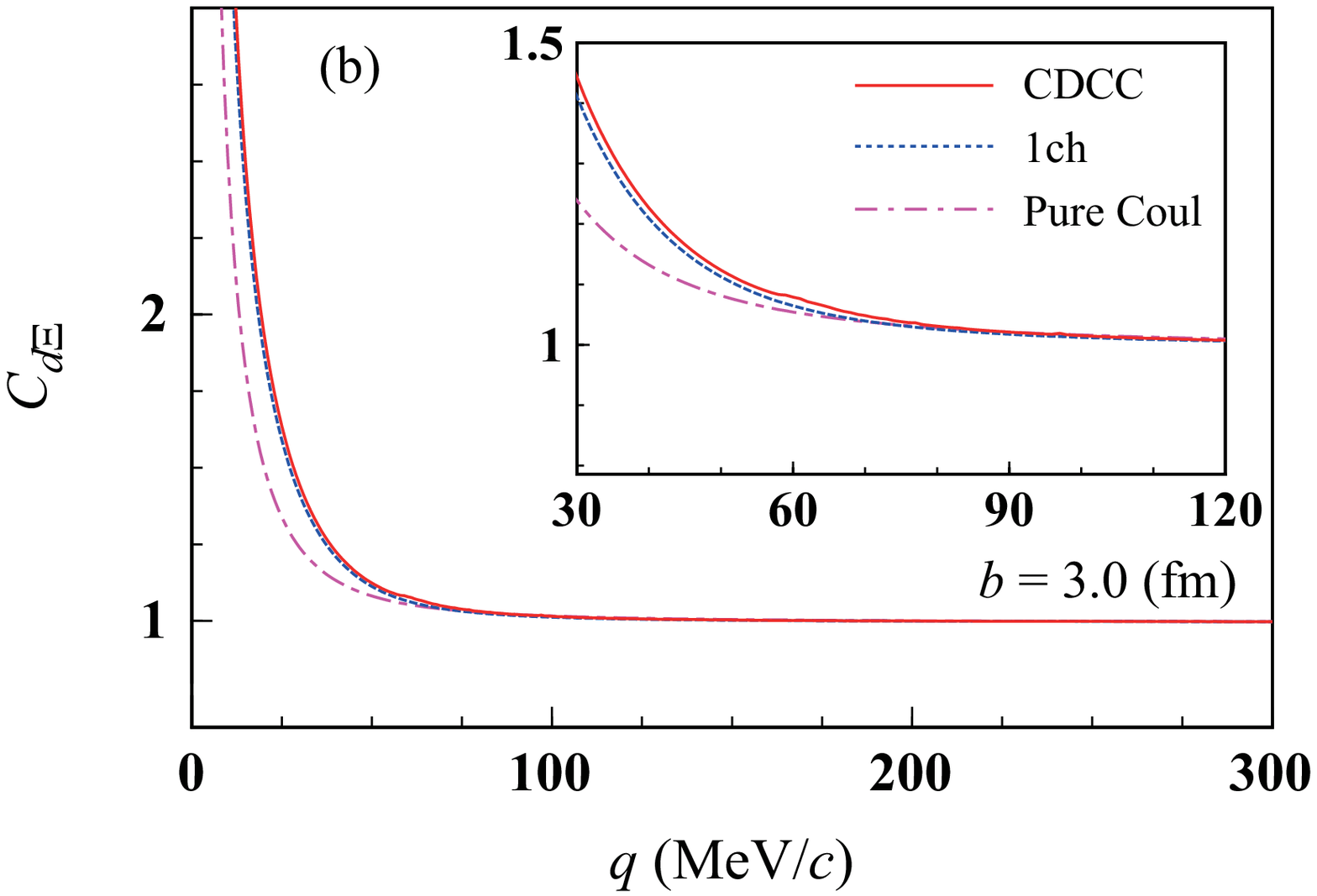}
\caption{Same as in Fig.~\ref{fig1} but with different values of the source size $b$ of the source function. (a) $b=1.6$~fm and (b) $b=3.0$~fm.}
\label{fig4}
\end{center}
\end{figure}
The convergence of the CDCC result regarding $\Delta_c$ for the $nn$ continuum is shown in Fig.~\ref{fig3}; $\Delta_c=0.2$~fm$^{-1}$ is used for the $pn$ continuum and $k_{\rm max}$ is set to 2.0~fm$^{-1}$ for both $pn$ and $nn$ continua. The solid (red), dashed (green), dotted (blue), and dash-dotted (purple) lines correspond to $\Delta_c=1.0$, 0.5, 0.2, and 0.005~fm$^{-1}$, respectively. The dash-dotted line is the same as the solid-line in Fig.~\ref{fig1}. The dashed line turns out to have a rather sharp peak around 93~MeV$/c$. This happens when the lowest (pseudo) $nn$ state is located just below the threshold energy; note that the eigenenergy of a discretized continuum state is defined by Eq.~(\ref{epsi}) and depends on $\Delta_c$. When $\Delta_c=0.2$~fm$^{-1}$, the eigenenergy of the lowest $nn$ state becomes 0.55~MeV and the peak appears at $q\sim65$~MeV$/c$. At the same time, another peak is found around 97~MeV$/c$, which corresponds to the second-lowest $nn$ state. As $\Delta_c$ becomes smaller, a larger number of peaks appear and the characteristics of each peak become less emphasized. It is found that with $\Delta_c=0.005$~fm$^{-1}$, a reasonably smooth $C_{d\Xi^-}$ is obtained. The shoulder structure of the dash-dotted line around 60~MeV$/c$ is due to many tiny peaks corresponding to low-lying $nn$ breakup states. It should be noted that for breakup states that do not strongly couple to the deuteron ground state, the above-mentioned threshold effect is negligibly small. This is why we can use a rather large bin-size, $\Delta_c=0.2$~fm$^{-1}$, for the $pn$ breakup states. The properties of the CC potentials for the $pn$ and $nn$ breakup states are discussed in Sec.~\ref{sec34}.

We show in Fig.~\ref{fig4} the dependence of $C_{d\Xi^-}$ on the source size $b$ of the source function; Fig.~\ref{fig4}(a) and Fig.~\ref{fig4}(b) correspond to $b=1.6$ and 3.0~fm, respectively. The meaning of each line is the same as in Fig.~\ref{fig1}. As $b$ increases, the correlation due to the strong interaction becomes weak, as well as the deuteron breakup effect. This is simply because the non $s$-wave contribution of the $d$-$\Xi$ scattering wave function is large in the outer region of $R$. Notwithstanding, the effect of the strong interaction on $C_{d\Xi^-}$ will remain at small $q$.

\subsection{Discretized continuum states of the $NN$ system}
\label{sec33}

%---- figure 5 ----%
\begin{figure}[H]
\begin{center}
\includegraphics[width=0.48\textwidth]{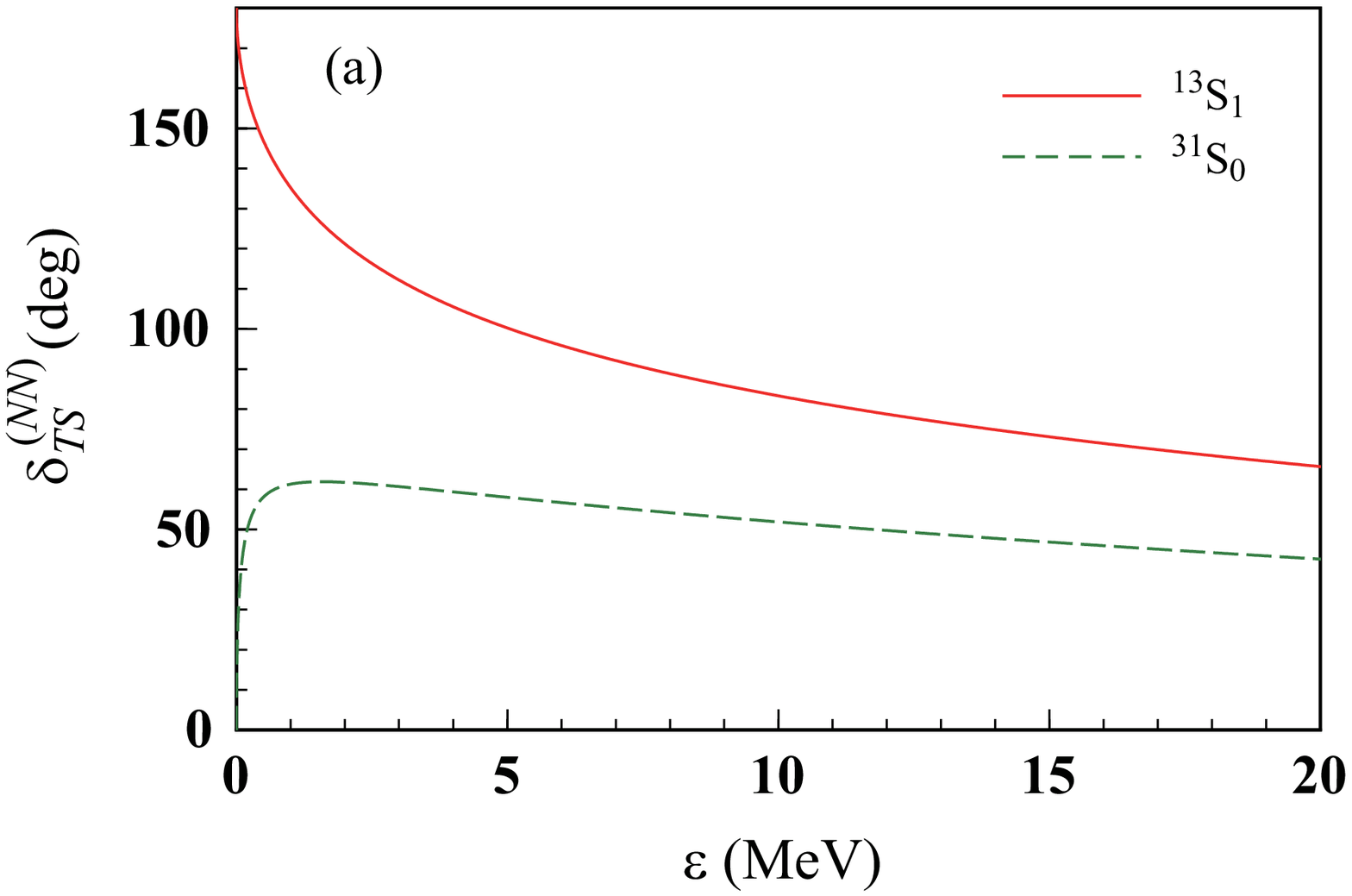}
\includegraphics[width=0.48\textwidth]{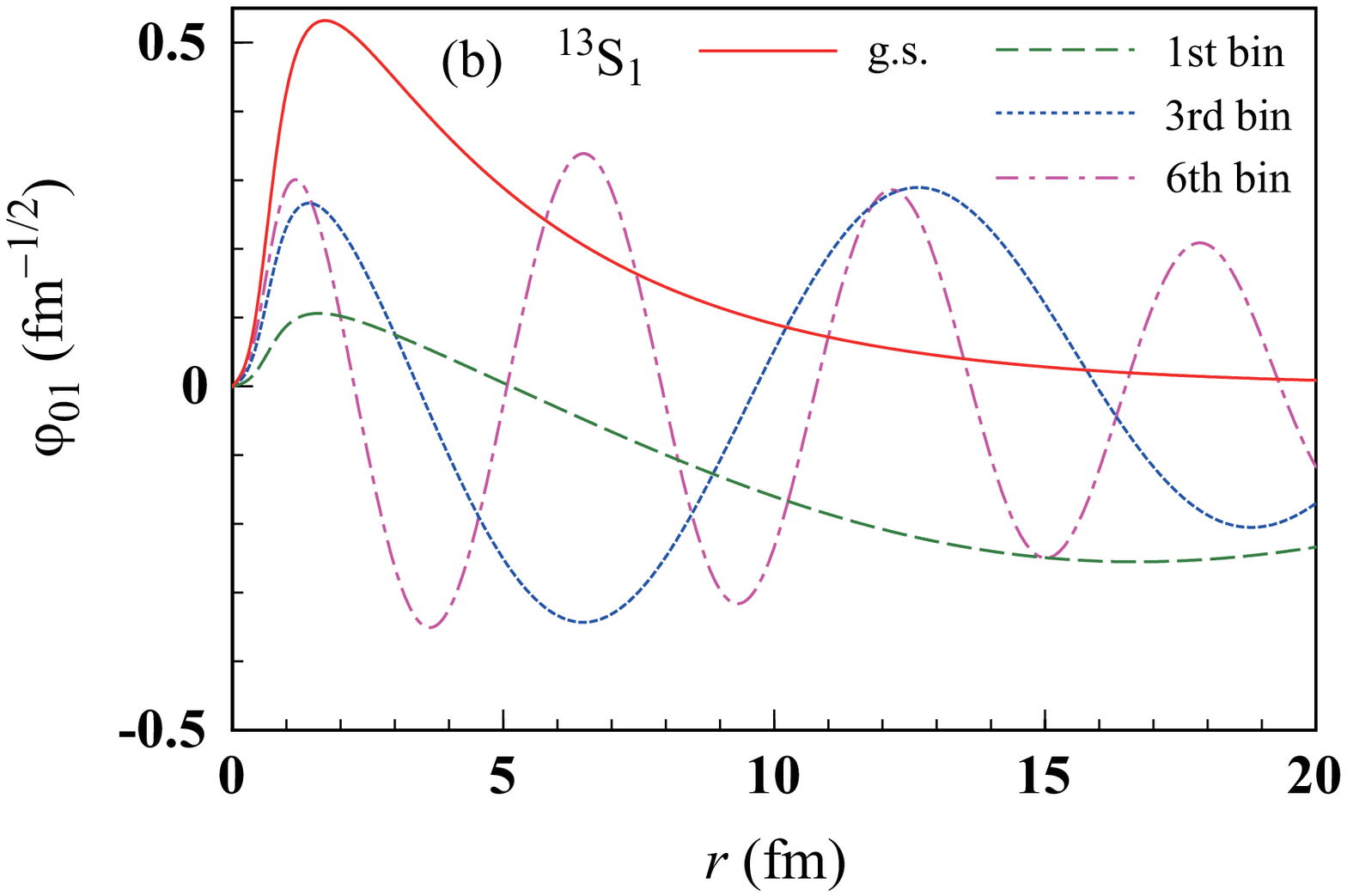}
\includegraphics[width=0.48\textwidth]{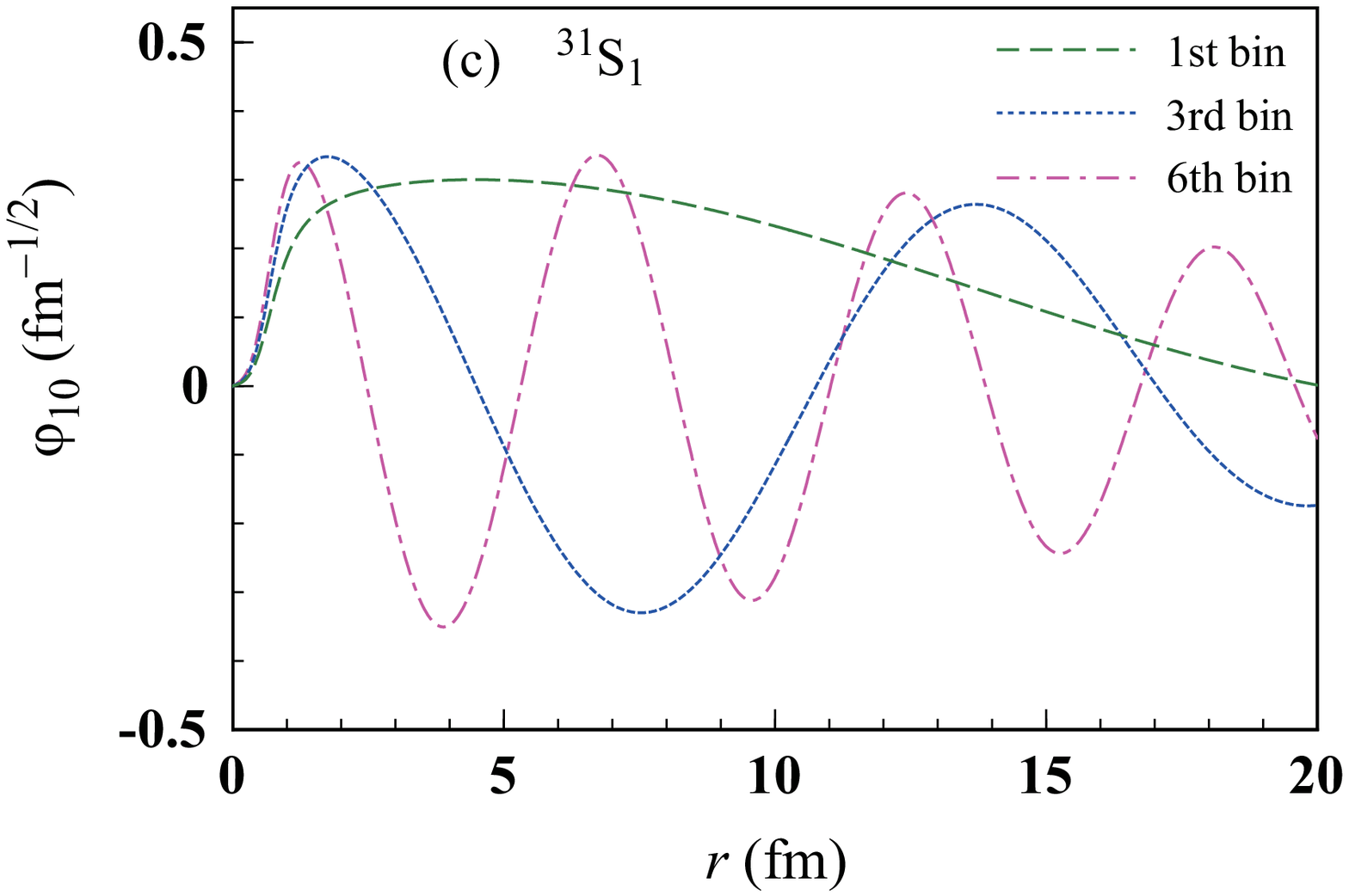}
\caption{(a) $NN$ $s$-wave scattering phase shift as a function of the c.m. scattering energy in the $^{13}$S$_1$ (solid line) and $^{31}$S$_0$ (dashed line) channels. (b) $NN$ discretized continuum states in the $^{13}$S$_1$ channel as a function of the distance of the two nucleons. The dashed, dotted, and dash-dotted lines correspond to the first, third, and sixth bin states, respectively, with the bin size of 0.2~fm$^{-1}$. The solid line represents the bound-state wave function of deuteron. (c) Same as in (b) but in the $^{31}$S$_0$ channel; there is no bound state in this channel.
}
\label{fig5}
\end{center}
\end{figure}
In this subsection, we discuss the properties of the $NN$ states included in the current study. For transparent discussion, the results below are evaluated with $\Delta_c=0.2$~fm$^{-1}$ ($\sim 40$~MeV/$c$) for both the $pn$ and $nn$ channels. As shown in Fig.~\ref{fig3}, apart from the threshold effect of the low-lying $nn$ breakup states, $C_{d\Xi^-}$ calculated with $\Delta_c=0.2$~fm$^{-1}$ well reproduces that with $\Delta_c=0.005$~fm$^{-1}$ ($\sim 1$~MeV/$c$). Therefore, discussion on $\varphi_{c}$ generated with $\Delta_c=0.2$~fm$^{-1}$ will be meaningful to understand the role of the $NN$ continuum in this study.

The $s$-wave phase shift $\delta^{(NN)}_{TS}$ of the $NN$ system is shown in Fig.~\ref{fig5}(a) as a function of the $NN$ c.m. energy $\varepsilon$. The solid (red) and dashed (green) lines represent $\delta^{(NN)}_{01}$ ($pn$ channel) and $\delta^{(NN)}_{10}$ ($nn$ channel), respectively. As is well known, $\delta^{(NN)}_{10}$ shows a rapid increase near $\varepsilon=0$, which is due to the virtual state (pole) of the $nn$ system. Although it is different from a resonance, the $nn$ wave function near the zero energy has a compact form as shown below. In Fig.~\ref{fig5}(b) and Fig.~\ref{fig5}(c), respectively, we show $\varphi_{i01}$ and $\varphi_{i10}$; in each panel, the dashed (green), dotted (blue), and dash-dotted (purple) lines correspond to the first bin ($k=0.0$--0.2~fm$^{-1}$, $\varepsilon_c=0.55$~MeV), the third bin ($k=0.4$--0.6~fm$^{-1}$, $\varepsilon_c=10.5$~MeV), and the sixth bin ($k=1.0$--1.2~fm$^{-1}$, $\varepsilon_c=50.3$~MeV) states, respectively. For comparison, the deuteron wave function is shown by the solid (red) line in Fig.~\ref{fig5}(b). As mentioned, the first bin state of the $nn$ channel behaves like a bound state. On the other hand, for the $pn$ channel, the amplitude of the first bin state in the inner region is very small, which makes this state almost decoupled from the deuteron ground state and other $NN$ states. As for the third bin state, the $pn$ wave function is slightly more shrunk than the $nn$ one, reflecting the difference in the phase shift shown in Fig.~\ref{fig5}(a). The dependence of the sixth bin state on the spin-isospin is found to be very small, which is the case also for higher bin states.

\subsection{$NN$-$\Xi$ coupled-channel potentials}
\label{sec34}

%---- figure 6 ----%
\begin{figure}[htbp]
\begin{center}
\includegraphics[width=0.48\textwidth]{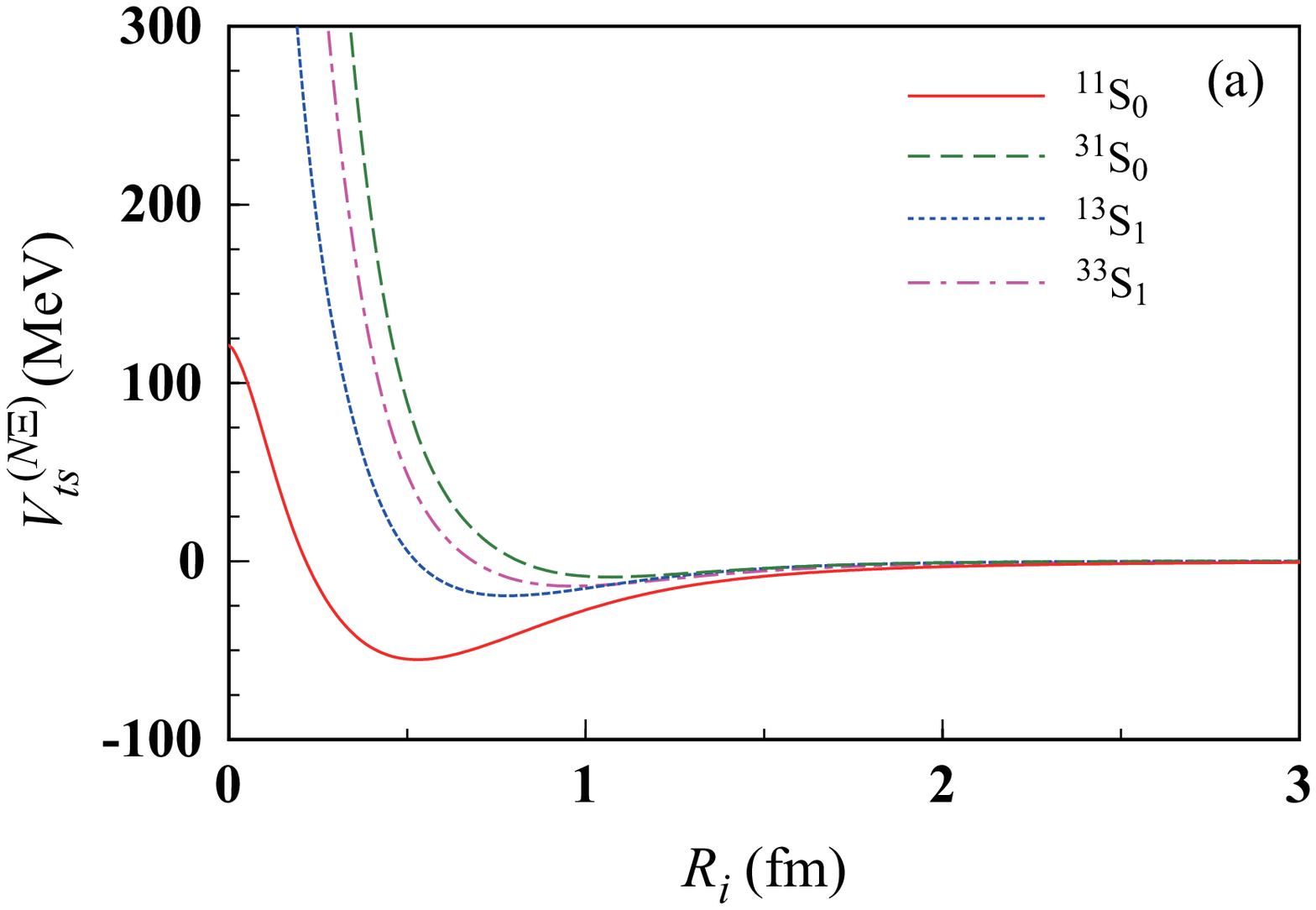}
\includegraphics[width=0.48\textwidth]{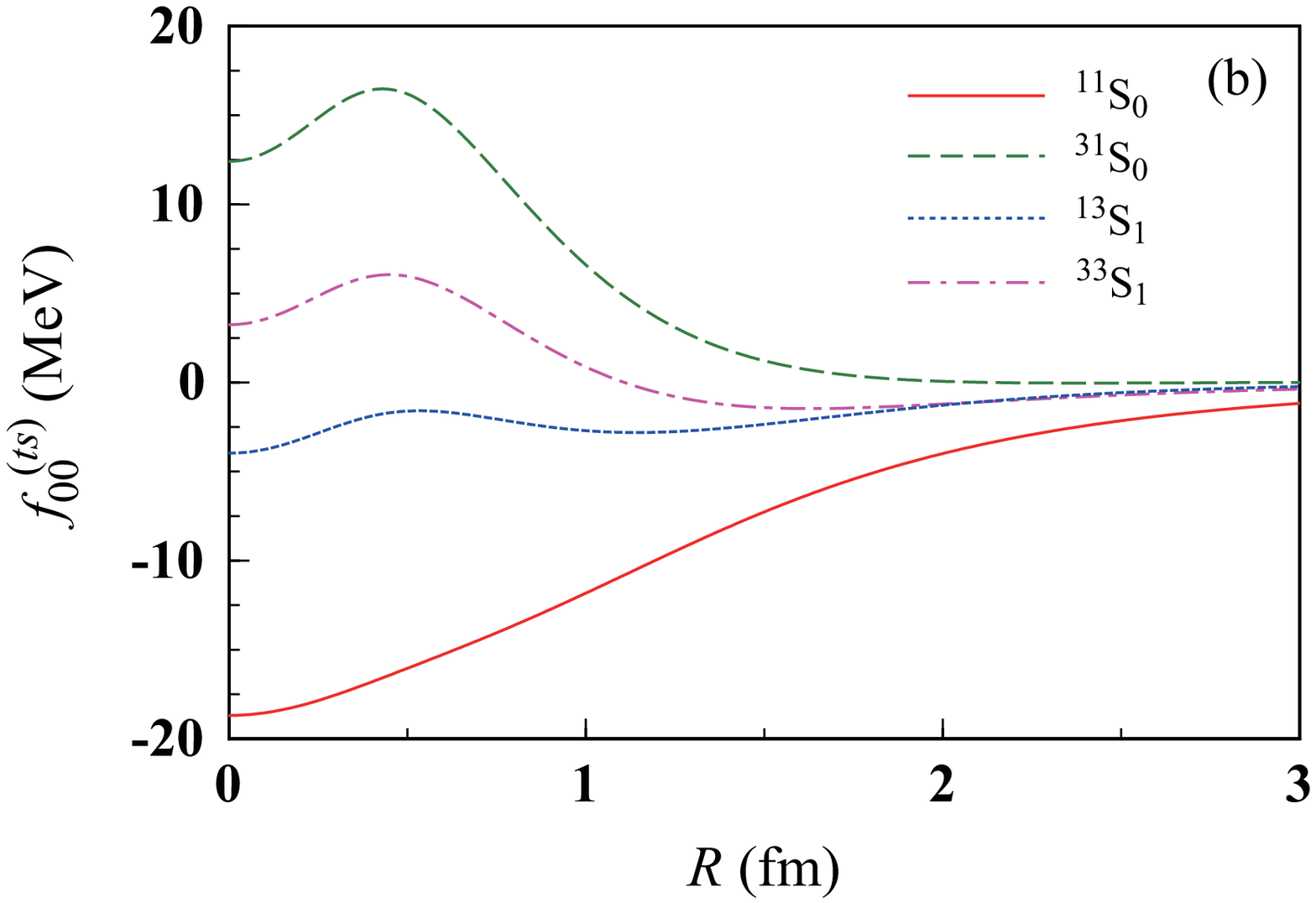}
\caption{(a) $N$-$\Xi$ interaction as a function of their distance. The solid, dashed, dotted, and dash-dotted lines correspond to the $^{11}$S$_0$, $^{31}$S$_0$, $^{13}$S$_1$, $^{33}$S$_1$ channels, respectively. (b) Same as in (a) but folded by the deuteron ground-state density.}
\label{fig6}
\end{center}
\end{figure}
The $N$-$\Xi$ interactions in individual spin-isospin channels as a function of the $N$-$\Xi$ distance are shown in Fig.~\ref{fig6}(a) and the corresponding folded potentials for the ground-ground channel, $f^{(ts)}_{00}$, are shown in Fig.~\ref{fig6}(b) as functions of $R$. In each panel, the potentials for $^{11}{\rm S}_0$, $^{31}{\rm S}_0$, $^{13}{\rm S}_1$, and $^{33}{\rm S}_1$ are represented by the solid (red), dashed (green), dotted (blue), and dash-dotted (purple) lines, respectively. Through the folding procedure, the characteristics of the potential for each channel becomes very clear. The potential in the $^{11}{\rm S}_0$ channel is attractive, while that in $^{31}{\rm S}_0$ repulsive. The feature of the potential in the $^{13}{\rm S}_1$ ($^{33}{\rm S}_1$) channel is similar to that in the $^{11}{\rm S}_0$ ($^{31}{\rm S}_0$) channel but with the absolute value weaken considerably. Note, however, that the attractive nature of the $N$-$\Xi$ potential in the $^{33}{\rm S}_1$ channel is found to remain when folded by the deuteron density; the $d$-$\Xi^-$ scattering length evaluated by taking only the $^{33}{\rm S}_1$ channel in the single-channel calculation is negative.
Here, we use the nuclear physics convention for the scattering length, that is,
\begin{equation}
\kappa \cot \delta
=
-\frac{1}{a_{s}}+\frac{r_s}{2} \kappa^2+O(\kappa^4),
\end{equation}
where $\kappa$ is the relative wave number of the two particles, $\delta$ is the $s$-wave scattering phase shift, $a_{s}$ is the scattering length, and $r_{s}$ is the effective range. The negative scattering length thus means that there is no bound state.
The qualitative features of $f^{(ts)}_{00}$ mentioned above are found to remain for other components of the folded potential.

From now on, we discuss the properties of the CC potentials $U_{cc'}^{\left(\sigma\right)}$. For simplicity, we take only four states of the $NN$ system, that is, the deuteron ground state ($d$), the third bin state in the $pn$ channel ($pn$), the first bin state in the $nn$ channel ($^2n$), and the third bin state in the $nn$ channel ($nn$); we abbreviate these four state as denoted in the parentheses. Here, as in Sec.~\ref{sec33}, we take $\Delta_c=0.2$~fm$^{-1}$ ($\sim 40$~MeV/$c$) for both channels.

%---- figure 7 ----%
\begin{figure}[htbp]
\begin{center}
\includegraphics[width=0.48\textwidth]{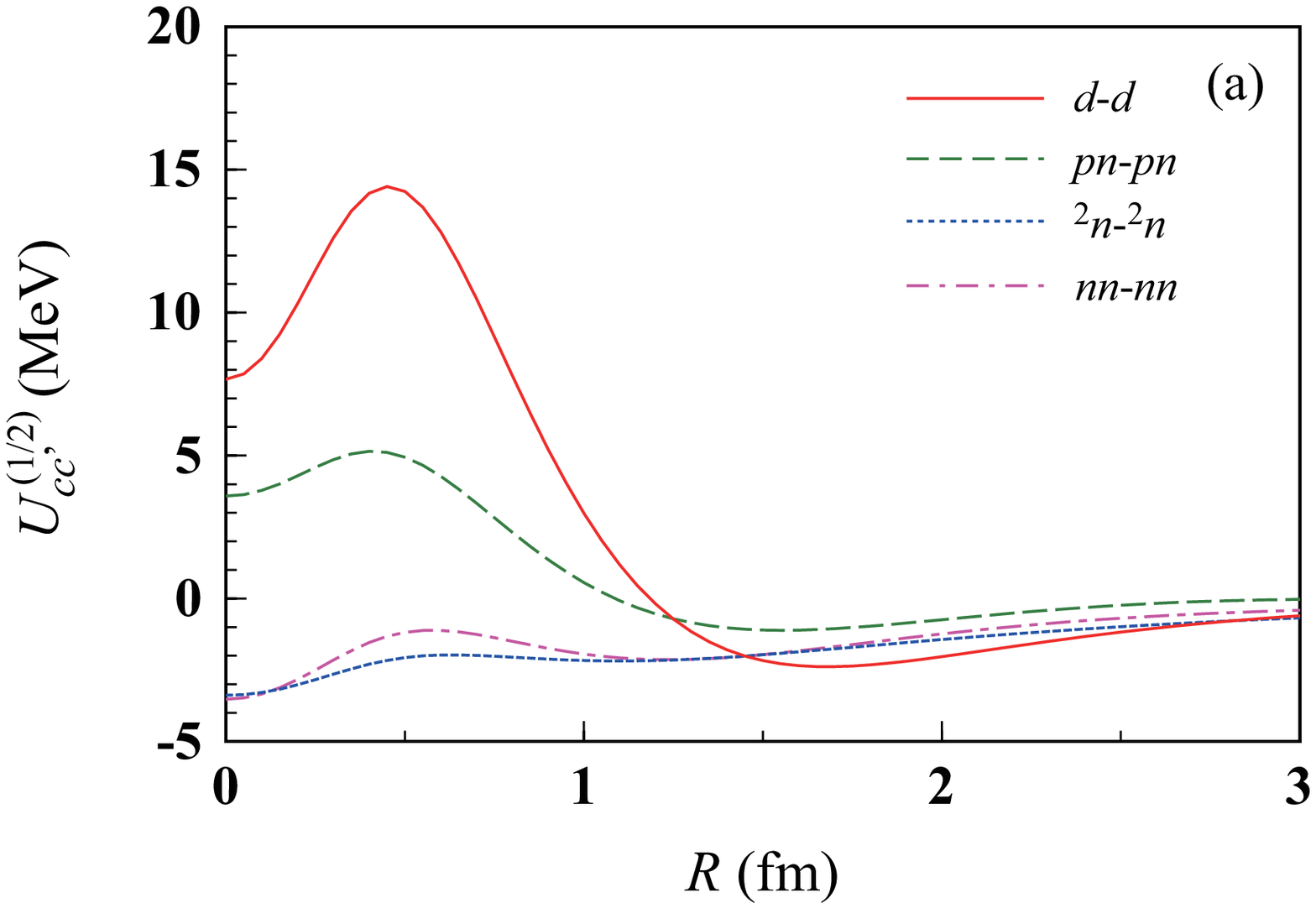}
\includegraphics[width=0.48\textwidth]{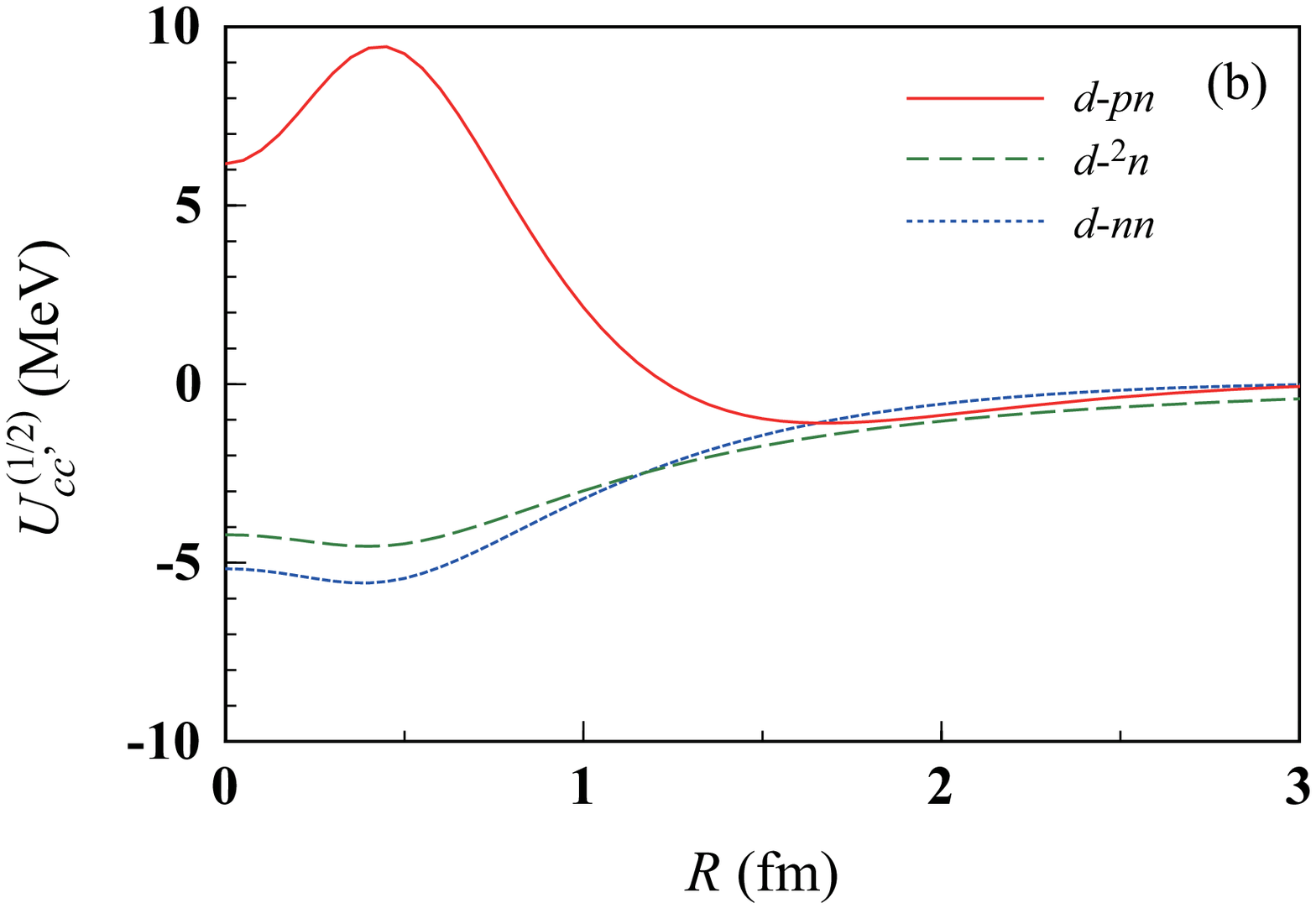}
\caption{Coupling potentials $U_{cc'}^{(1/2)}(R)$. (a) Diagonal components for the deuteron ground state ($d$), the third bin state in the $^{13}$S$_1$ channel ($pn$). the first bin state in the $^{31}$S$_0$ channel (${^2}n$), and the third bin state in the $^{31}$S$_0$ channel ($nn$) are represented by the solid, dashed, dotted, and dash-dotted lines, respectively. (b) Potentials for the $d$-$pn$ (solid line), $d$-$^{2}n$ (dashed line), and $d$-$nn$ couplings.}
\label{fig7}
\end{center}
\end{figure}
In Fig.~\ref{fig7}(a), we show the diagonal part of the CC potentials for the four states; the total channel spin $\sigma$ is taken to be $1/2$. The solid (red), dashed (green), dotted (blue), and dash-dotted (purple) lines correspond to the $d$, $pn$, $^2n$, and $nn$ states, respectively. The former two are repulsive in the interior region ($R \la 1$~fm) and weakly attractive at larger $R$, whereas the latter two are attractive in the entire region. This is due to the spin-isospin selection given by Eqs.~(\ref{ucctete}) and (\ref{uccsese}) combined with the spin-isospin dependence of the folded potential shown in Fig.~\ref{fig6}(b). The result shown in Fig.~\ref{fig7}(a) indicates that an $nn$ pair can be closer to the $\Xi$ particle than a $pn$ pair including deuteron. This is one of the main reasons for the significant breakup effect from the $nn$ channel.

Figure~\ref{fig7}(b) represents the coupling potential between the deuteron channel and each of the other three; the solid (red), dashed (green), and dotted (blue) lines correspond to the $d$-$pn$, $d$-$^2n$, and $d$-$nn$ couplings, respectively. It should be noted that the sign of non-diagonal coupling potentials has no meaning. As mentioned above, the folded potential $f^{(ts)}_{cc'}$ does not strongly depend on the combination of the channels, $c$ and $c'$. Consequently, the qualitative feature of the $d$-$pn$ coupling potential is similar to that of the $d$-$d$ diagonal potential. The behavior of the $d$-$^2n$ and $d$-$nn$ couplings can be understood by Eq.~(\ref{ucctese}) and Fig.~\ref{fig6}(b). An important remark is that the magnitude of the $d$-$^2n$ coupling potential is comparable to the $d$-$nn$ and $d$-$pn$ ones because of the compactness of the $^2n$ wave function as shown in Fig.~\ref{fig5}(c) (the dashed line). This feature is also crucial for making the breakup effect of the $nn$ channel important. Note that the coupling between the deuteron ground state and a low-lying $pn$ state is significantly weaker than the results shown in Fig.~\ref{fig7}(b).

To complete the discussion on the breakup effect, we need to consider the scattering threshold effect as well. When the $d$-$\Xi^-$ c.m. scattering energy $E_0$ is low, the channel energy $E_c$ for the $^2n$ channel becomes negative. In this case, even though the $d$-$^2n$ coupling is strong and the $^2n$-$^2n$ diagonal potential is attractive, the scattering wave $\chi_c^{(\sigma)}$ has to be considerably quenched because of the damping boundary condition of Eq.~(\ref{bcc}). An exception occurs when $E_c$ is very close to the threshold, that is, $E_c\sim 0$. This is how the shoulder structure of $C_{d\Xi^-}$ is developed (see also Sec.~\ref{sec35}). 

%---- figure 8 ----%
\begin{figure}[htbp]
\begin{center}
\includegraphics[width=0.48\textwidth]{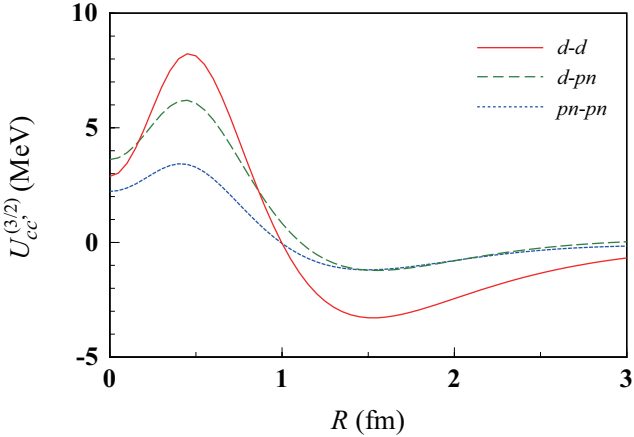}
\caption{The $d$-$d$ diagonal, $d$-$pn$ coupling, and $pn$-$pn$ diagonal potentials with $\sigma=3/2$ are shown by the solid, dashed, and dotted lines, respectively.}
\label{fig8}
\end{center}
\end{figure}
We show in Fig.~\ref{fig8} the coupling potentials with $\sigma=3/2$, for which the $nn$ states are not allowed. The solid (red), dashed (green), and dotted (blue) lines show the $d$-$d$ diagonal, $d$-$pn$ coupling, and $pn$-$pn$ diagonal potentials, respectively. The features of the results can be understood by Eq.~(\ref{ucc3h}) and Fig.~\ref{fig6}(b). It is found that the absence of the $nn$ channel makes the breakup effect negligibly small when $\sigma=3/2$, as shown in Sec.~\ref{sec35}.

%---- figure 9 ----%
\begin{figure}[htbp]
\begin{center}
\includegraphics[width=0.48\textwidth]{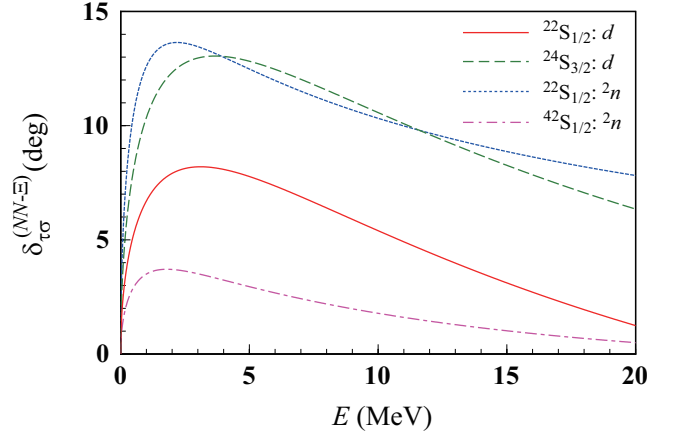}
\caption{$s$-wave phase shift of the $NN$-$\Xi$ scattering as a function of the c.m. scattering energy. The solid and dashed lines correspond to the $d$-$\Xi^-$ scattering in the $^{22}$S$_{1/2}$ and $^{24}$S$_{3/2}$ channels, respectively. The $nn$-$\Xi$ scattering phase shift in the $^{22}$S$_{1/2}$ ($^{22}$S$_{1/2}$) channel is shown by the dotted (dash-dotted) line; the discretized $nn$ continuum state corresponding to the wave number between 0.0~fm$^{-1}$ and 0.2~fm$^{-1}$ is adopted as an internal $nn$ wave function. All the results are obtained by the single-channel calculation.
}
\label{fig9}
\end{center}
\end{figure}
Figure~\ref{fig9} displays the nuclear scattering phase shift $\delta^{(NN\mbox{-}\Xi)}_{\tau\sigma}$ of the $NN$-$\Xi$ system as a function of the c.m. scattering energy. The solid (red) and dashed (green) lines show $\delta^{(NN\mbox{-}\Xi)}_{1/2,1/2}$ and $\delta^{(NN\mbox{-}\Xi)}_{1/2,3/2}$, respectively. In the calculation, a single-channel scattering problem with $U_{00}^{\left(\sigma\right)}\left(R\right)$ is solved for $\sigma=1/2$ and 3/2. As shown in Fig.~\ref{fig9}, the net effect of $U_{00}^{\left(\sigma\right)}\left(R\right)$ is found to be attractive, and the attraction of $U_{00}^{\left(3/2\right)}\left(R\right)$ is stronger than that of $U_{00}^{\left(1/2\right)}\left(R\right)$. In Table~\ref{tab1}, we show the $s$-wave scattering length $a_{s}$ and the effective range $r_{s}$ for the $NN$-$\Xi$ scattering.
\begin{table}[htbp]
\caption{$s$-wave scattering lengths $a_{s}$ and the effective ranges $r_{s}$ for the $NN$-$\Xi$ scattering calculated by the single-channel calculation.
}
\label{tab1}
\begin{center}
\begin{tabular}{cccccccc}
\hline\hline
 $T$ & $S$ & $\varepsilon_c$ (MeV) & $k$ (fm$^{-1}$)& $\tau$ & $\sigma$ & $a_s$ (fm) & $r_s$ (fm) \\
\hline
  0  &  1  & $-2.25$  & ------   & 1/2 & 1/2 & $-0.7164$  & 14.4 \\
  0  &  1  & $-2.25$  & ------   & 1/2 & 3/2 & $-1.1073$  & 9.21 \\
  1  &  0  & 0.553  & 0.0--0.2   & 1/2 & 1/2 & $-2.8629$  & 4.02 \\
  1  &  0  & 0.553  & 0.0--0.2   & 3/2 & 1/2 & $-0.57851$ & 16.1 \\
\hline\hline
\end{tabular}
\end{center}
\end{table}

In Fig.~\ref{fig9}, we also show the results of the phase shift by the $^2n$-$^2n$ diagonal potential for $(\tau,\sigma)=(1/2,1/2)$ and $(3/2,1/2)$ by the dotted (blue) and dash-dotted (purple) lines, respectively. The behavior of the dotted line is similar to that of the dashed line, indicating a rather strong attraction of the $^2n$-$^2n$ potential for the $(\tau,\sigma)=(1/2,1/2)$ channel as $U_{00}^{\left(3/2\right)}\left(R\right)$. In the $(\tau,\sigma)=(3/2,1/2)$ channel, which is irrelevant to the $d$-$\Xi^-$ scattering, the attraction of the $^2n$-$^2n$ diagonal potential is found to be weak. The values of $a_s$ and $r_s$ by the $^2n$-$^2n$ potential are also shown in Table~\ref{tab1}. One should be careful, however, that the results shown in Fig~\ref{fig9} and Table~\ref{tab1} regarding the $^2n$-$^2n$ diagonal potential depend on the definition of the $^2n$ state. In the current discussion, we regard the discretized continuum state corresponding to $k=0.0$--0.2~fm$^{-1}$ as the $^2n$ state. Because we here adopt a single-channel calculation, those results will easily change if we adopt a different bin-size for the $^2n$ state. Investigation of the $^2n$-$\Xi^0$ scattering within the framework of CDCC will be an interesting subject, but it is beyond the scope of this study. Notwithstanding, the results obtained by the $^2n$-$^2n$ diagonal potential in the current definition will be helpful to understand the qualitative features of the breakup effects on $C_{d\Xi^-}$ through $nn$ low-lying continuum states.

\subsection{$NN$-$\Xi$ scattering wave functions}
\label{sec35}

In this subsection, we see the $NN$-$\Xi$ scattering wave functions $\chi_c^{(\sigma)}$ as a result of the CC effect discussed so far. We adopt the numerical setting written in Sec.~\ref{sec31} with which a converged result of $C_{d\Xi^-}$ is obtained. The source size of the source function is taken to be 1.2~fm. We choose three values of $q$; $q=30$, 60, and 100~MeV$/c$.
These values are selected regarding the $nn$-$\Xi^0$ threshold momentum of about 60~MeV/$c$ in this study. However, this is due to the neglect of the isospin dependence of the particle masses. In reality, the $nn$-$\Xi^0$ threshold lies 3~MeV below the $d\Xi^-$ threshold and the $nn$-$\Xi^0$ channel is open for all values of $q$ discussed so far. Notwithstanding, we will discuss the behavior of $\chi_c^{(\sigma)}$ at below, near, and above the $nn$-$\Xi^0$ threshold energy corresponding to the model adopted in this study. In all the figures shown below, contributions from the $pn$ continuum states are not shown because these are negligibly small. We also omit the discussion of the $\sigma=3/2$ channel because of the negligibly small CC effect.

%---- figure 10 ----%
\begin{figure}[htbp]
\begin{center}
\includegraphics[width=0.48\textwidth]{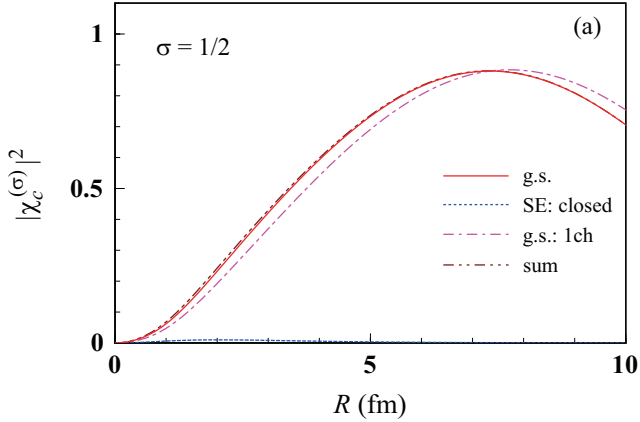}
\caption{Absolute square of the $NN$-$\Xi$ scattering wave function for $\sigma=1/2$ at $q=30$~MeV/$c$. The solid line represents the elastic channel component, whereas the dotted line is the sum of the components for the $nn$ closed channel; both are obtained by CDCC. The dash-dot-dotted line shows the total contributions of the channels included. The dash-dotted line is the result of the single-channel calculation.
}
\label{fig10}
\end{center}
\end{figure}
We show in Fig.~\ref{fig10} the result with $\sigma=1/2$ at $q=30$~MeV$/c$, in which all the breakup states are closed. The solid (red) line shows the contribution of the $d$-$\Xi^-$ elastic-channel component, that is, $|\chi_0^{(\sigma)}|^2$, whereas the dotted (blue) line shows the sum of $|\chi_c^{(\sigma)}|^2$ of the $nn$ states in the closed channels. The dash-dot-dotted (brown) line is the sum of the contributions from all the channels. For comparison, we show by the dash-dotted (purple) line $|\chi_0^{(\sigma)}|^2$ obtained with the single-channel calculation; it is denoted by $|\chi_0^{(\sigma){\rm 1ch}}|^2$ below. One can see that the contribution of the $nn$ breakup states is very small, whereas $|\chi_0^{(\sigma)}|^2$ is somewhat larger than $|\chi_0^{(\sigma){\rm 1ch}}|^2$. This is the source of the enhancement of $C_{d\Xi^-}$ due to the deuteron breakup. It indicates that the coupling through the breakup states acts as an additional attractive potential for the $d$-$\Xi^-$ elastic channel. It is found that the $pn$ breakup states are also responsible for this back-coupling to the elastic channel, though their importance is considerably less than that of the $nn$ states, as seen from Fig.~\ref{fig1}.

%---- figure 11 ----%
\begin{figure}[htbp]
\begin{center}
\includegraphics[width=0.48\textwidth]{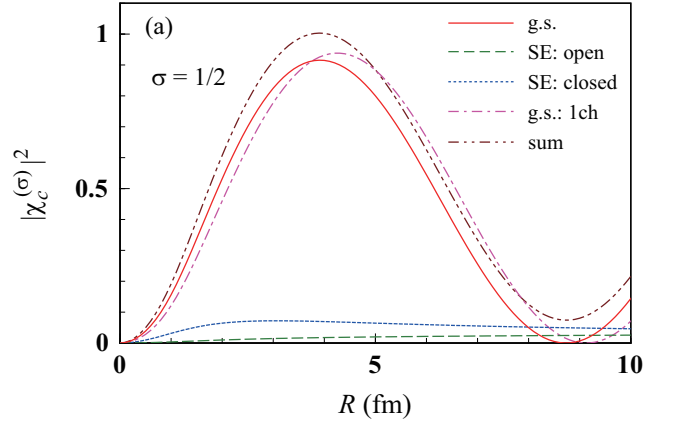}
\caption{Same as in Fig.~\ref{fig10} but at $q=60$~MeV/$c$. The dashed line shows the sum of the contributions from the $nn$ open breakup channels.}
\label{fig11}
\end{center}
\end{figure}

The results at $q=60$~MeV$/c$ are shown in Fig.~\ref{fig11}. The meaning of the lines is the same as in Fig.~\ref{fig10} but the dashed (green) line  shows the contribution of the $nn$ channels for which $E_c>0$ (open channels). One sees the back-coupling effect on $|\chi_0^{(1/2)}|^2$ as at 30~MeV$/c$. On top of that, the contribution of the closed $nn$ channel is appreciable (the dotted line). As a result, the difference between $|\chi_0^{(\sigma){\rm 1ch}}|^2$ and the sum of $|\chi_c^{(1/2)}|^2$ is more developed than at 30~MeV$/c$. The reason for this enhancement is that the channel energies of the closed $nn$ channels are close to zero. One sees that the dotted (blue) line in Fig.~\ref{fig11} decreases very slowly at large $R$.

%---- figure 12 ----%
\begin{figure}[htbp]
\begin{center}
\includegraphics[width=0.48\textwidth]{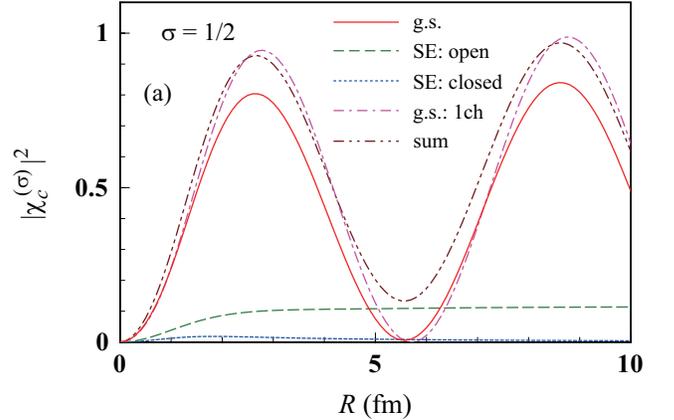}
\caption{Same as in Fig.~\ref{fig11} but at $q=100$~MeV/$c$.}
\label{fig12}
\end{center}
\end{figure}
At $q=100$~MeV$/c$, quite a lot of channels become open. As shown by the dashed (green) line in Fig.~\ref{fig12}, the contribution of the open $nn$ breakup channels becomes important. However, the magnitude of the sum of all the channels (the dash-dot-dotted line) is very similar to that of $|\chi_0^{(\sigma){\rm 1ch}}|^2$ (the dash-dotted line). This is because of the unitarity of the scattering matrix, that is, the conservation of the flux. This feature makes the net breakup effect on $C_{d\Xi^-}$ very small, though a slight enhancement at small $R$ remains. It will be worth pointing out that the three-body wave functions for the $nn$ open breakup channels may contribute to $C_{d\Xi^-}$ in a different manner if we use a more sophisticated source function. This will be another important subject in future.

\section{summary}
\label{sec4}

We have evaluated for the first time the $d$-$\Xi^-$ correlation function $C_{d\Xi^-}$ with a three-body reaction model including the $s$-wave breakup states of deuteron (both $pn$ and $nn$ continua). The continuum-discretized coupled-channels method (CDCC) is adopted to describe the $N+N+\Xi$ three-body wave function for the $d$-$\Xi^-$ scattering. The Argonne V4' $NN$ force and a parametrization of the $N$-$\Xi$ interaction by the LQCD method are employed in the three-body model calculation. We have assumed that only the $s$-wave scattering wave between the c.m. of the $NN$ system and $\Xi$ is affected by the strong interaction and the Coulomb interaction between $d$ and $\Xi^-$ is approximated to be present in all the isospin channels. A simplified source function independent of channels and $NN$ relative coordinate is employed, and the isospin dependence of the masses of $N$ and $\Xi$ are disregarded. A clear enhancement of $C_{d\Xi^-}$ due to the strong interaction is confirmed as in preceding studies.

We have found that $C_{d\Xi^-}$ increases by the deuteron breakup effect by 6--8~\% at the $d\hyphen\Xi^-$ relative momentum $q$ below 70~MeV/$c$. This is mainly due to the back-coupling to the elastic channel through the low-lying $nn$ continuum, the tail of the $nn$ virtual state. The key mechanism of this enhancement is that the low-lying $nn$ continuum wave function is spatially compact and the spin-isospin selection makes the $nn$-$\Xi^0$ potential attractive in the entire region. Besides, when the c.m. scattering energy is close to the $nn$-$\Xi^0$ threshold, the $nn$-$\Xi^0$ channel component in the total three-body wave function itself becomes important. Consequently, a shoulder structure of $C_{d\Xi^-}$ is developed around $q=60$~MeV/$c$, though in reality, the $nn$-$\Xi^0$ threshold is located below $q=0$. At larger $q$, although the deuteron breakup probability becomes larger, the unitarity condition on the scattering matrix makes the net breakup effect on $C_{d\Xi^-}$ very limited.

Because the deuteron breakup effect on $C_{d\Xi^-}$ is not very significant, the finding of this study may justify the studies on $C_{d\Xi^-}$ by including only the deuteron ground state, except for the additional enhancement of $C_{d\Xi^-}$ by several percent. It will be important, however, to investigate the deuteron breakup effect with a more realistic three-body source function. There will be a possibility to access the $n+n+\Xi^0$ state in the relativistic heavy-ion collision through $C_{d\Xi^-}$. Direct detection of multi-neutron as done in low-energy nuclear physics will be even more interesting. On the theory side, modification on the treatment of the Coulomb interaction in isospin-dependent three-body scattering will be necessary.
At the same time, the mass difference between $\Xi^-$ and $\Xi^0$
amounts to be around 7~MeV and needs to be taken care of.
Together with the mass difference of $p$ and $n$ and the deuteron binding energy,
the $nn$-$\Xi^0$ threshold lies 3~MeV below the $d$-$\Xi^-$ threshold
and the effects of the dineutron state $^2n$ may be more important.

%---- acknowledgements ----%
\section*{ACKNOWLEDGEMENTS}
This work has been supported in part by the Grants-in-Aid for Scientific Research from JSPS (No. JP19H01898, No. JP19H05151, and No. JP21H00125), by the Yukawa International Program for Quark-hadron Sciences (YIPQS), by the National Natural Science Foundation of China (NSFC) under Grant No.~11835015 and No.~12047503, by the NSFC and the Deutsche Forschungsgemeinschaft (DFG, German Research Foundation) through the funds provided to the Sino-German Collaborative Research Center TRR110 {\lq\lq}Symmetries and the Emergence of Structure in QCD'' (NSFC Grant No.~12070131001, DFG Project-ID 196253076), by the Chinese Academy of Sciences (CAS) under Grant No.~XDB34030000 and No.~QYZDB-SSW-SYS013, the CAS President's International Fellowship Initiative (PIFI) under Grant No.~2020PM0020, and China Postdoctoral Science Foundation under Grant No.~2020M680687.

\appendix

\section{Three-body wave function with incoming boundary condition}
\label{a1}
In the evaluation of the correlation function, we need a scattering wave function corresponding to the incoming boundary condition, that is, the time-reversed solution $\Psi_{M_0\mu_0}^{(-)}$. To obtain it, we first rewrite Eq.~(\ref{totwf}) as
\begin{eqnarray}
\Psi_{M_0\mu_0}^{(+)}(r,R)
&=&
\sum_{T'S'M'\mu'}
\Theta_{T'}^{\left(\frac{1}{2},-\frac{1}{2} \right)}
\eta_{S' M'}^{\left(NN\right)}
\eta_{\frac{1}{2} \mu'}^{\left(\Xi\right)} \nonumber \\
&&\times
\bar{\Psi}_{S'M'\mu';1M_0\mu_0}^{(+)}(r,R)
\end{eqnarray}
with
\begin{eqnarray}
\bar{\Psi}_{S'M'\mu';1M_0\mu_0}^{(+)}(r,R)
&=&
\sqrt{4\pi}\sum_{\sigma m_{\sigma}}
\left(1 M_0\frac{1}{2}\mu_0\Big|\sigma m_{\sigma}\right)
e^{i\sigma_{0}} \nonumber\\
& & \times\sum_{i'}
\frac{\varphi_{c'}\left(r\right)}{r}
\frac{\chi_{c'}^{(\sigma)}\left( K_{c'},R\right)}{K_0R}
\frac{1}{4\pi} \nonumber\\
& & \times
\left(S' M'\frac{1}{2}\mu'\big|\sigma m_{\sigma}\right).
\nonumber\\
\label{totwf2}
\end{eqnarray}
Then $\Psi_{M_0\mu_0}^{(-)}$ is given by
\begin{eqnarray}
\Psi_{M_0\mu_0}^{(-)}(r,R)
&=&
\sum_{T'S'M'\mu'}
\Theta_{T'}^{\left(\frac{1}{2},-\frac{1}{2} \right)}
\eta_{S' M'}^{\left(NN\right)}
\eta_{\frac{1}{2} \mu'}^{\left(\Xi\right)} \nonumber \\
&&\times
\bar{\Psi}_{S'M'\mu';1M_0\mu_0}^{(-)}(r,R)
\end{eqnarray}
with
\begin{eqnarray}
\bar{\Psi}_{S'M'\mu';1M_0\mu_0}^{(-)}(r,R)
&=&
(-)^{1+M_0+\mu_0 - S'-M'-\mu'} \nonumber \\
&&\times
\bar{\Psi}_{S',-M',-\mu';1,-M_0,-\mu_0}^{(+)*}(r,R)
\nonumber \\
&=&
(-)^{M_0+\mu_0-M'-\mu'} \nonumber \\
&&\times
\bar{\Psi}_{S'M'\mu';1M_0\mu_0}^{(+)*}(r,R).
\end{eqnarray}

\section{Monopole component of Gaussian}

When $V_{ts}^{(N\Xi)}$ has a Gaussian form
\begin{equation}
V_{ts}^{(N\Xi)}(R_i)=
\sum_j \bar{V}_{ts,j}^{(N\Xi)} e^{-\alpha_{ts,j} R_i^2},
\end{equation}
its monopole component is given by
\begin{equation}
V_{ts;0}^{(N\Xi)}(R,r)
=
\bar{V}_{ts,j}^{(N\Xi)}
\frac{e^{-\alpha_{ts,j}(R-r/2)^2}-e^{-\alpha_{ts,j}(R+r/2)^2}}{2\alpha_{ts,j}Rr}
\label{monog}
\end{equation}
for both $i=1$ and 2.

%---- reference ----%

\end{document}